\documentclass[twocolumn]{aastex63} 
\usepackage{graphicx}              

\received{Mar 03, 2022}
\revised{Apr 26, 2022}
\accepted{May 10, 2022}
\submitjournal{\apj}

\shorttitle{Gamma-ray Diagnostics of Neutron-Star Merger Remnants}
\shortauthors{Terada et al.}

\begin{document}

\title{Gamma-ray Diagnostics of r-process Nucleosynthesis in the Remnants of Galactic Binary Neutron-Star Mergers}

\correspondingauthor{Yukikatsu Terada}
\email{terada@mail.saitama-u.ac.jp}

\author[0000-0002-2359-1857]{Yukikatsu Terada}
\affiliation{Graduate School of Science and Engineering, Saitama University, 255 Sakura-ku, Saitama-shi, Saitama 338-8570, Japan}
\affiliation{Institute of Space and Astronautical Science, Japan Aerospace Exploration Agency, Sagamihara, Kanagawa, Japan}

\author{Yuya Miwa}
\affiliation{Graduate School of Science and Engineering, Saitama University, 255 Sakura-ku, Saitama-shi, Saitama 338-8570, Japan}

\author{Hayato Ohsumi}
\affiliation{Graduate School of Science and Engineering, Saitama University, 255 Sakura-ku, Saitama-shi, Saitama 338-8570, Japan}

\author{Shin-ichiro Fujimoto}
\affiliation{National Institute of Technology Kumamoto College, Kumamoto 861-1102, Japan}

\author{Satoru Katsuda}
\affiliation{Graduate School of Science and Engineering, Saitama University, 255 Sakura-ku, Saitama-shi, Saitama 338-8570, Japan}

\author{Aya Bamba}
\affiliation{Department of Physics, The University of Tokyo, 7-3-1 Hongo, Bunkyo, Tokyo 113-0033, Japan}
\affiliation{Research Center for the Early Universe, School of Science, The University of Tokyo, 7-3-1 Hongo, Bunkyo, Tokyo 113-0033, Japan}

\author{Ryo Yamazaki}
\affiliation{Department of Physical Sciences, Aoyama Gakuin University, 5-10-1 Fuchinobe, Sagamihara, Kanagawa 252-5258, Japan}
\affiliation{Institute of Laser Engineering, Osaka University, 2-6, Yamadaoka,Suita, Osaka 565-0871, Japan}

\begin{abstract}
We perform a full nuclear-network numerical calculation of the $r$-process nuclei in binary neutron-star mergers (NSMs), with the aim of estimating $\gamma$-ray emissions from the remnants of Galactic NSMs up to $10^6$ years old.
The nucleosynthesis calculation of 4,070 nuclei is adopted to provide the elemental composition ratios of nuclei with an electron fraction $Y_{\rm e}$ between 0.10 and 0.45 .
The decay processes of 3,237 unstable nuclei are simulated to extract the $\gamma$-ray spectra.
As a result, the NSMs have different spectral color in $\gamma$-ray band from various other astronomical objects at less than $10^5$ years old. 
In addition, we propose a new line-diagnostic method for $Y_{\rm e}$ that uses the line ratios of either $^{137{\rm m}}$Ba/$^{85}$K or $^{243}$Am/$^{60{\rm m}}$Co, which become larger than unity for young and old $r$-process sites, respectively, with a low $Y_{\rm e}$ environment.
From an estimation of the distance limit for $\gamma$-ray observations as a function of the age, the high sensitivity in the sub-MeV band, at approximately $10^{-9}$ photons s$^{-1}$ cm$^{-2}$ or $10^{-15}$ erg s$^{-1}$ cm$^{-2}$, is required to cover all the NSM remnants in our Galaxy if we assume that the population of NSMs by \citet{2019ApJ...880...23W}. 
A $\gamma$-ray survey with sensitivities of $10^{-8}$--$10^{-7}$ photons s$^{-1}$ cm$^{-2}$ or $10^{-14}$--$10^{-13}$ erg s$^{-1}$ cm$^{-2}$ in the 70--4000 keV band is expected to find emissions from at least one NSM remnant under the assumption of NSM rate of 30 Myr$^{-1}$.
The feasibility of $\gamma$-ray missions to observe Galactic NSMs are also studied.
\end{abstract}


\section{Introduction} 
\label{sec:intro}

Elements heavier than Bi exist in our universe, but their origin remains a mystery.
Most cosmic isotopes heavier than the iron group are expected to be created 
by the rapid-neutron capture process, also known as the {\it r}-process \citep{1957PASP...69..201C,1957RvMP...29..547B,1991PhR...208..267C,2006NuPhA.777..676W,2007PhR...442..237Q,2007PhR...450...97A}, 
but the actual nucleosynthesis sites 
capable of achieving such neutron-rich environments remains a matter of debate. 
Before the discovery of binary neutron-star mergers (NSM) observed as gravitational-wave objects like GW170817 \citep{2017PhRvL.119p1101A},
NSMs were considered to be more promising as {\it r}-process nucleosynthesis sites than other primary candidates, such as core-collapse supernovae (SNe), because NSMs could achieve more neutron-rich (lower electron fraction $Y_{\rm e}$) environments \citep{2011ApJ...726L..15W,1974ApJ...192L.145L,2010MNRAS.406.2650M}. 
The event rate of NSMs is much lower than that of SNe, 
but the yield of $r$-process nuclei in one event is expected to be very high \citep{2015NatCo...6.5956W,2015NatPh..11.1042H}.
Observational evidence of the existence of $r$-process nuclei has already been obtained 
by infrared observations of  kilonovae (also called macronovae or $r$-process novae)
in some short gamma-ray bursts, such as GRB~130603B \citep{2013Natur.500..547T} and 
the gravitational wave event GW170817 \citep{2017ApJ...851L..21V}.
However, the infrared radiation from NSMs is, in principle, the result of indirect emissions from unstable $r$-process nuclei, and any hint of elements heavier than the lanthanoids is still missing from the infrared information.
Given that the nuclear levels of nuclei are in the MeV energy range, the $\gamma$ rays from $r$-process nuclei should be the best probe for searching for $r$-process sites in the universe.

According to theoretical estimates of the $\gamma$-ray flux 
from binary NSMs \citep{2016MNRAS.459...35H}, 
the $\gamma$-ray radiation immediately following a merging event 
is very dim at about $10^{-8}$--$10^{-7}$ photons s$^{-1}$ cm$^{-2}$ keV$^{-1}$, even at an extremely close distance $d$ of 3 Mpc.
This flux is comparable with or below what the sensitivities of current and near-future MeV missions can detect.
The precise measurements of photon energies are, in principle, rather difficult in the MeV band, 
where Compton scattering dominates over the photon-absorption process. 
Therefore, the ability to detect $\gamma$ rays from NSMs by an immediate follow-up observation 
(a Target-of-opportunity observation; ToO) would be limited by the sensitivity of the $\gamma$-ray instruments.
Instead, a non-ToO observation of $\gamma$ rays from long-lived nuclei in NSMs 
would be an alternative way to survey $r$-process sites, and this has been proposed by \citet{2019ApJ...880...23W} \textcolor{red}{and \citet{2020ApJ...903L...3W}}.
The $\gamma$-ray luminosity from nuclei with long lifetimes, on the order of $10^3$--$10^6$ years, 
becomes much lower than that from short-lived nuclei, but 
if we limit the survey area within our Galaxy ($d \lesssim$ 10 kpc), 
then the $\gamma$-ray flux in non-ToO observations is expected to become comparable with what is required for ToO observations.
Therefore, non-ToO observations should provide more sensitive $\gamma$-ray surveys of NSMs,
because the exposure time (the accumulation time of signals) is not limited like it is in ToO observations.
Another benefit from performing a non-ToO survey is the better identification of $\gamma$-ray lines; 
we expect the effect of Doppler broadening to be smaller for older NSM remnants than for very young NSMs.

Here we focus on the non-ToO survey of $\gamma$ rays from $r$-process nuclei 
in a possible Galactic NSM remnant.
In this paper, we estimate $\gamma$-ray emissions from Galactic NSM remnants 
in an older age range than in previous work \citep{2016MNRAS.459...35H,2020ApJ...903L...3W} by 
using nuclear-network numerical calculations with a complete nuclear database.
This paper also aims to provide $\gamma$-ray diagnostic methods for NSMs, 
showing the required sensitivities for future $\gamma$-ray observatories.
In our study, we assume that the $\gamma$-ray instruments have a wider field-of-view (FOV) 
than the object size of the NSM remnants, which are larger than early NSMs in a ToO observation.
We also assume that the instruments accumulate all of the $\gamma$-ray emissions from the NSM remnants,
even though the nuclei may mix with the circumstellar medium (CSM) during the evolution of the remnants.

The rest of this paper is organized as follows.
In Section \ref{sec:estimation}, 
we summarize our environments and procedures for the nuclear-network numerical calculation and 
show the results for $\gamma$-ray emissions from NSM remnants.
In Section \ref{sec:diagnostic},
we present the $\gamma$-ray diagnostics that utilize spectral color to identify NSM remnants  
and provide the line properties for estimating the age $t$ and $Y_{\rm e}$.
In Section \ref{sec:discussion},
we discuss the survey distance and coverage in our Galaxy permitted by the instrument sensitivities,
the corresponding limitation of the NSM rate in our Galaxy, and expectations for future missions.

\section{Numerical estimation of Gamma Rays from Neutron-Star Merger Remnants}
\label{sec:estimation}
\subsection{Overview of Numerical Calculation}
\label{sec:estimation:summary}
To estimate the $\gamma$-ray emissions from binary NSM remnants of various ages, 
we performed a numerical simulation comprising the following three steps: 
{\bf 1)} calculation of the mass distribution of $r$-process nuclei for NSMs at $t=1$ year,
{\bf 2)} calculation of the decay processes of unstable nuclei emitting $\gamma$ rays,
and
{\bf 3)} a simple calculation of the radiation transfer of $\gamma$ rays from NSMs.

\begin{figure}[ht]
    \centering
    \includegraphics[width=0.45 \textwidth]{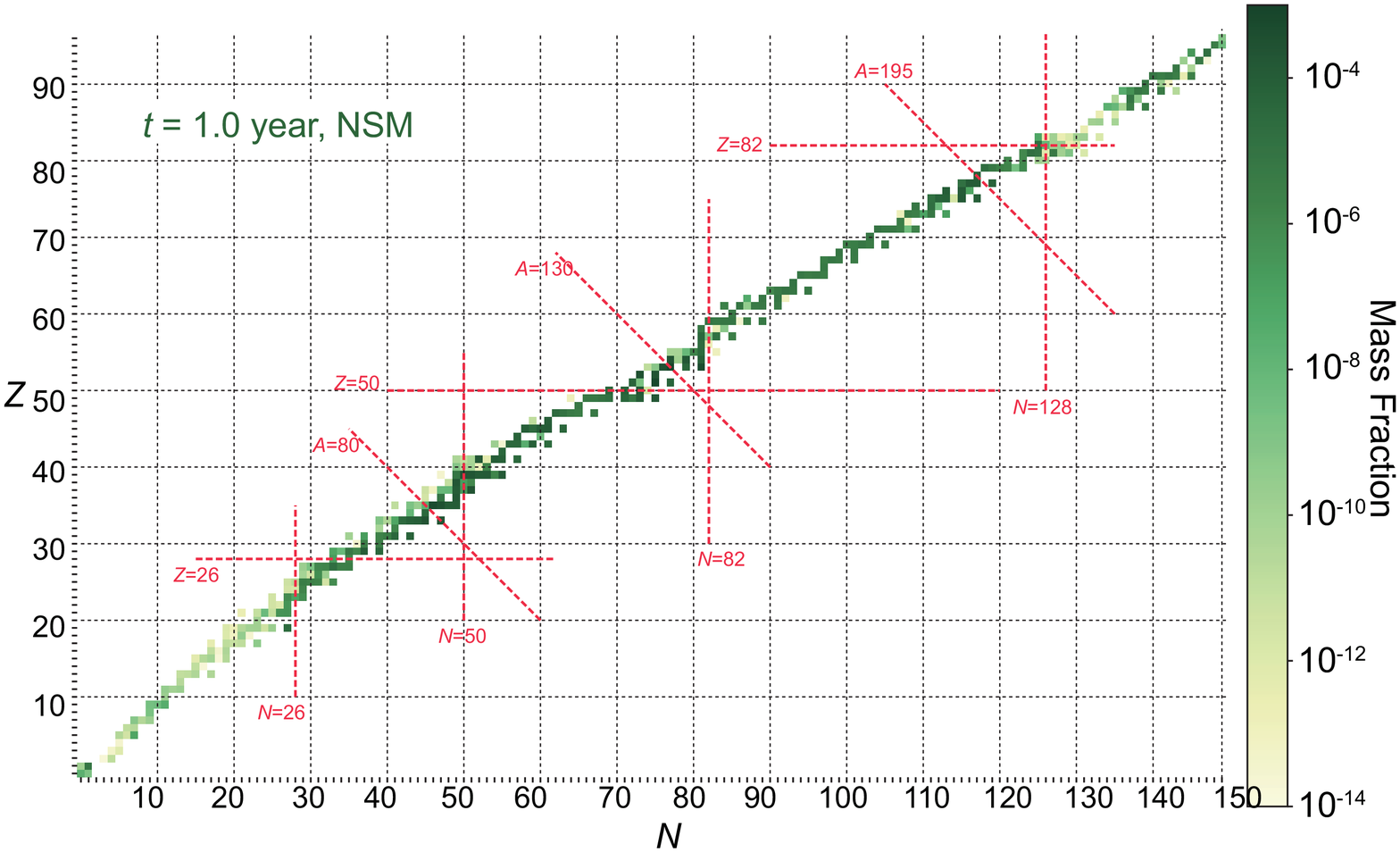}
    \caption{The r-process nuclei in the NSM case at $t=1.0$ year on the table of nuclides.
    The magic numbers for neutrons and protons are indicated by red dashed lines}
    \label{fig:initial_tableOfNuclides}
\end{figure}

\begin{figure}[ht]
    \centering
    \includegraphics[width=0.45 \textwidth]{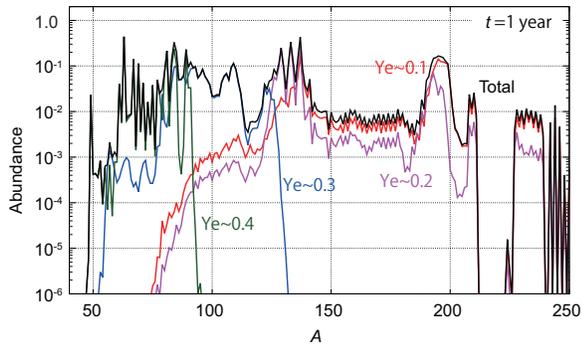}
    \caption{Mass number distribution of the abundance for the NSM case at $t=1.0$ year, with $Y_{\rm e} \sim $ 0.1, 0.2, 0.3, and 0.4, shown in red, magenta, blue, and green, respectively.}
    \label{fig:initial_massDistribution}
\end{figure}

For the first step, we adopted the nucleosynthesis calculation for around 4,070 nuclei performed by \citet{2007ApJ...656..382F}, 
which was cooled using the adiabatic expansion modeled from \citet{1999ApJ...525L.121F} 
to provide the elemental composition ratios of nuclei
for $Y_{\rm e} = 0.10, 0.15, 0.20, 0.25, 0.30, 0.35, 0.40$, and $0.45$.
This estimation assumes that the initial environment has 
a temperature of $9\times 10^9$ K, radius of 100 km, 
entropy per baryon of 10$k_{\rm B}$, where $k_{\rm B}$ is the Boltzmann constant, 
and  velocity of $2\times 10^9$ cm s$^{-1}$,
along with the initial abundances of the 4,070 nuclei in nuclear statistical equilibrium.
As a result, the calculation provides the mass fractions at t = 1 year 
evaluated with the nuclear reaction network (network A in \citealt{2007ApJ...656..382F}),
by using $Y_{\rm e} = 0.10$--$0.45$ in steps of 0.05.
To set up the mass distribution of nuclei for the NSMs at $t=1$ year, 
we blended the nuclei with the mass fraction using the $Y_{\rm e}$ provided in \citet{2014ApJ...789L..39W}.
Specifically, the fractions are 4.54\%, 4.85\%, 14.6\%, 29.7\%, 10.3\%, 25.1\%, 10.5\%, and 0.33\% for $Y_{\rm e}=0.10, 0.15, 0.20, 0.25, 0.30, 0.35, 0.40$, and $0.45$, respectively.
Note that this $Y_{\rm e}$-fraction model by \citet{2014ApJ...789L..39W} describes slightly-less neutron-rich environment than those by the recent dynamical-ejecta models after the kilonova observations of the gravitational event GW170817, such as four models in \citet{2022MNRAS.510.2804K} under two kinds of equation-of-states, density dependent 2 (DD2) \citep{2010NuPhA.837..210H,2010PhRvC..81a5803T} and SFHo \citep{2013ApJ...774...17S}.
In this paper, we adopted the first one by \citet{2014ApJ...789L..39W} as a pessimistic case for the $r$-process site, but changing the $Y_{\rm e}$-fraction models does not change the conclusions from the $\gamma$-ray spectra as tested in the later section \ref{sec:diagnostic}.
Figure \ref{fig:initial_tableOfNuclides} shows the mass fraction of multiple nuclei at $t=1$ year generated in an NSM case, information that is given in the table of nuclides (neutron number $N$ vs atomic number $Z$).
Using the same data set, Figure \ref{fig:initial_massDistribution} summarizes 
the distribution of nuclei with mass number $A$ at $t=1$ year, 
showing the contributions of $Y_{\rm e}$.
This plot demonstrates that the environment with lower $Y_{\rm e}$ contributes to the generation of heavier elements.

For the second step, we simulated the decay processes of unstable nuclei, 
starting from the mass distribution at $t=1$ year calculated in the first step.
We used the Decay Data File 2015 (DDF-2015) \citep{2016JENDL_DDF2015_a}
in the Japanese Evaluated Nuclear Data Library (JENDL) \citep{2016JENDL_DDF2015_b},
which provides the decay profiles of 3,237 nuclei up to Z=104 (Rf).
Here we applied a correction to the $\gamma$-ray information for $^{241}$Am; 
this error was reported from our study and was fixed in the next version of the database.
The originality of this study lies in the comprehensiveness of nuclei treated in the calculation. 
In the nuclear-decay calculation, we adopted the 
$\alpha$-decay, $\beta^{-}$-decay, $\beta^{+}$-decay, electron capture, 
isomeric transition, and $\gamma$-decay processes.
In our calculation, the internal conversion process is ignored, 
which emits soft X-rays and makes a negligible contribution to the $\gamma$-ray band.
The neutron- and proton-emission processes are also ignored 
because they contribute to the very early phase, which is out of the scope of this study.
The spontaneous fission process may occur on $^{257}$Es, $^{256}$Cf, $^{254}$Cf, and $^{250}$Cm,
but its contribution is negligible. Thus, this process is also excluded from our calculation.
In addition, we do not calculate the $\gamma$-ray emission from secondary electrons (electrons from $\beta$-decay, $\delta$-rays, and so on) after the decay of unstable nuclei.

\begin{figure}[ht]
    \centering
    \includegraphics[width=0.45 \textwidth]{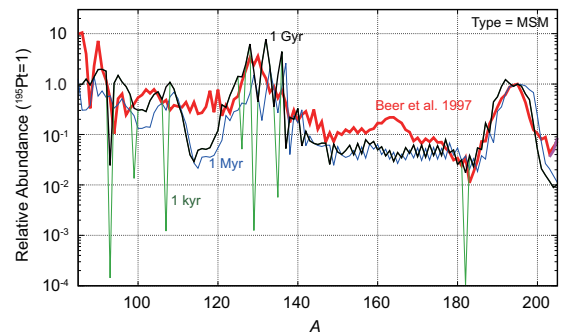}
    \caption{The relative abundance of $^{195}$Pt in the NSM case at $t=$ 1 thousand, 1 million, and 1 billion years, shown in green, blue, and thick black, respectively, compared with the semi-empirical abundance distribution of \citet{1997ApJ...474..843B}.}
    \label{fig:mass_timeevolve_long}
\end{figure}

To verify the calculations in the second step, we refer to 
Figure \ref{fig:mass_timeevolve_long}, which represents the relative abundances of nuclei at $t=$ 1 thousand, 1 million,
and 1 billion years as a function of $A$.
The distribution in $A$ does not change dramatically after $t\sim$ 1 thousand years,
except for the disappearance of the small dips at the magic numbers.
The distribution for $t> 1$ million years becomes roughly consistent 
with the semi-empirical abundance distribution of cosmic $r$-process nuclei 
in \citet{1997ApJ...474..843B}, which is close to the solar abundance distribution.

For the outputs of the second step,
we obtain the $\gamma$-ray flux $F_{\gamma, i}$ from the NSM at distance $d$.
Using the nuclear $\gamma$-ray intensity $I_{\gamma, i}$ of the $i$-th element with
the mass number $A_{i}$, the mass $M_{i}$, and the half-life $T_{1/2}$, 
the $F_{\gamma, i}$ in a small time interval $dt$ is described as
\begin{equation}
    F_{\gamma, i} = \frac{N_{\rm A}}{4 \pi d} \frac{M_i}{A_i} \frac{I_{\gamma, i}}{T_{1/2}} 
    \left(\frac{1}{2}\right)^{-dt/T_{1/2}},
\end{equation}
where $N_{\rm A}$ is Avogadro's number.

Finally, the third step is to calculate the transfer of $\gamma$ rays through the NSM ejecta. 
However, we omitted the detailed Monte-Carlo calculation of the radiation transfer
because the optical depth decreases rapidly after the merger event, 
by roughly $\propto \sqrt{t}$ \citep{2019ApJ...872...19L}; within the scope of our study at $t \gg 1$ year, the optical depth is thin and negligible. 
Therefore, the degradation of the line profiles by Compton scattering is not included in our calculation,
which would be dominant in only the very early phase.
Note that the detail calculations of the MeV $\gamma$-ray spectra from Galactic NSMs in the initial phase were performed by \citet{2020ApJ...903L...3W,2021ApJ...923..219W}.
In this step, 
we apply only the bulk Doppler-broadening effect caused by the expansion velocity $v(t)$.
The thermal Doppler-broadening effect is ignored in this calculation
because it is two or three orders of magnitude smaller than that from the expansion motion
of the heavy elements in the $A=$ 50--200 range.
In reality, the line profile from the bulk Doppler effect becomes complicated due to the complex contributions of various velocity components, as has been observed in the X-ray lines from heavy elements in SN remnants \citep{2017ApJ...834...19G,2018PASJ...70...88K}. 
For simplicity, we applied the Gaussian distribution function for the line profile in the calculation.
Of the various velocity elements in the remnant, we applied only single Gaussian broadening to the maximum velocity component, which we assume to be the forward shock motion. This was done to simulate the most robust case for considering the $\gamma$-ray sensitivity.
In our assumption, 
$v(t)$ starts from the initial value $v(0)=0.3$ $c$, where $c$ is the speed of light, 
and evolves at a constant rate during the free expansion phase.
During the Sedov-Taylor phase, $v(t)$ evolves as $v(t) \propto t^{-(3/5)}$ \citep{1950RSPSA.201..159T}, 
and then as $v(t) \propto t^{-(0.7)}$ during the pressure-driven snowplow (PDS) phase \citep{1977ApJ...218..148M}.
We assume that the free expansion, Sedov-Taylor, and PDF phases end
at $t=$ 10, $4.7 \times 10^4$, and $1.65 \times 10^6$ years, respectively. 
The ages of these phase transitions may change by about one order of magnitude due to differences in density of the CSM, but this modification affects only the Doppler-broadening effect. It becomes negligible when compared with the typical energy resolutions of $\gamma$-ray instruments for ages older than $t\sim 1,000$ years, the range that lies within the scope of this study.
Note that even at $t=10^6$ years, $v(t)$ approaches $\sim 20$ km s$^{-1}$, which is about double the speed of sound for a typical CSM density of 0.01 cm$^{-3}$. The radius becomes $\sim 100$ pc.
Finally, we get the $\gamma$-ray spectra for NSMs at $t$,
accumulated from all of the $r$-process nuclei in the ejecta.

\subsection{Gamma-ray Emission and Evolution}
\label{sec:estimation:gamma}

\begin{figure*}[ht]
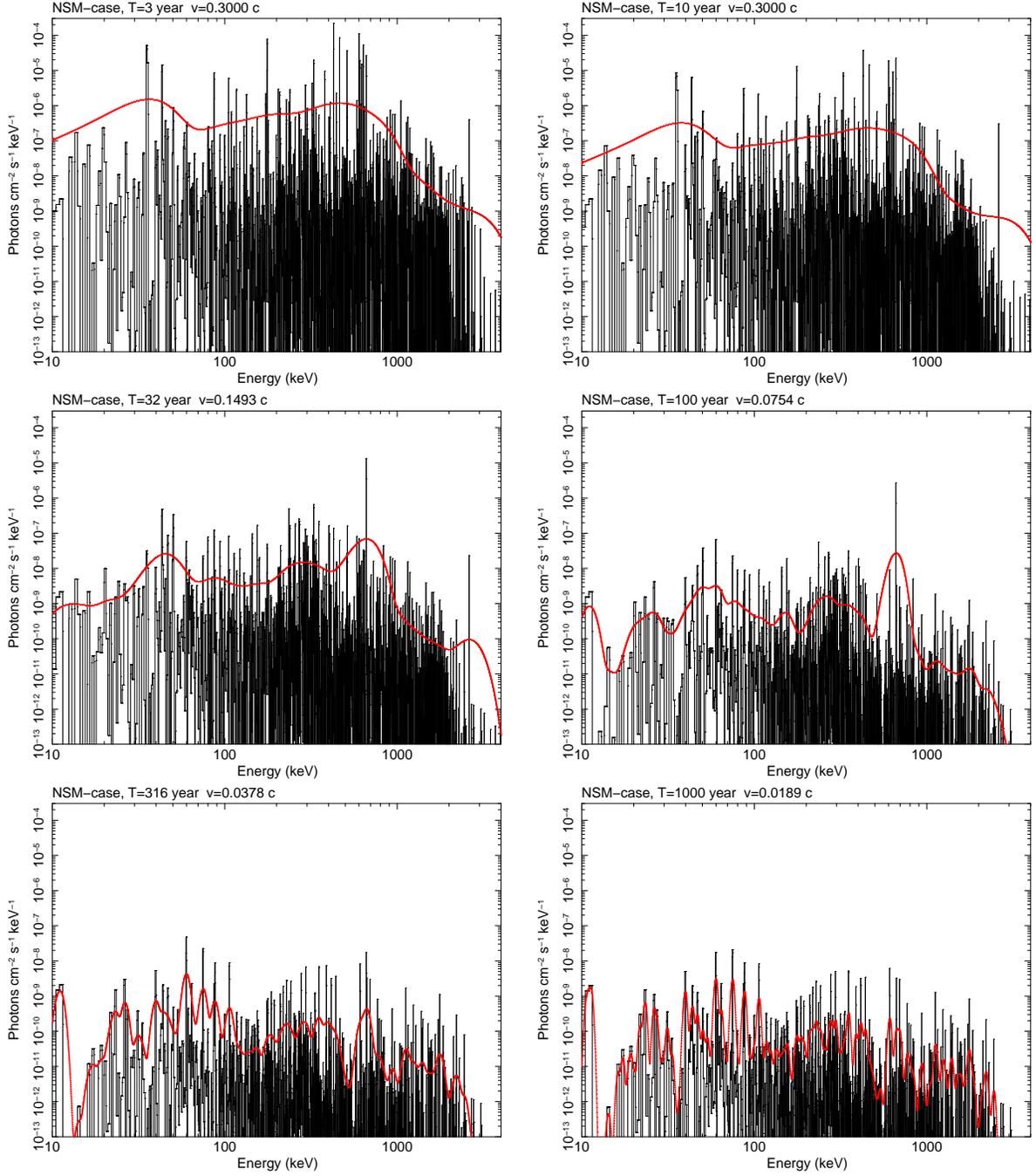

    \begin{center}
    \includegraphics[width=0.33 \textwidth, angle=-90]{gammasim_yterada_apj_figure04_3yr.eps}
    \includegraphics[width=0.33 \textwidth, angle=-90]{gammasim_yterada_apj_figure04_10yr.eps}
    \includegraphics[width=0.33 \textwidth, angle=-90]{gammasim_yterada_apj_figure04_32yr.eps}
    \includegraphics[width=0.33 \textwidth, angle=-90]{gammasim_yterada_apj_figure04_100yr.eps}
    \includegraphics[width=0.33 \textwidth, angle=-90]{gammasim_yterada_apj_figure04_316yr.eps}
    \includegraphics[width=0.33 \textwidth, angle=-90]{gammasim_yterada_apj_figure04_1000yr.eps}
    \end{center}
    \caption{Gamma-ray spectra simulated for the NSM case at $t=$ 3 - 1,000 years, assuming a distance of 10 kpc with an initial velocity of 0.3 $c$ (see the text). The red and black plots represent the spectra with and without the Doppler-broadening effect, respectively. The time since the merging event and the velocities are shown in each top label.}
    \label{fig:rprocess_g_model_plot1}
\end{figure*}

\begin{figure*}[ht]
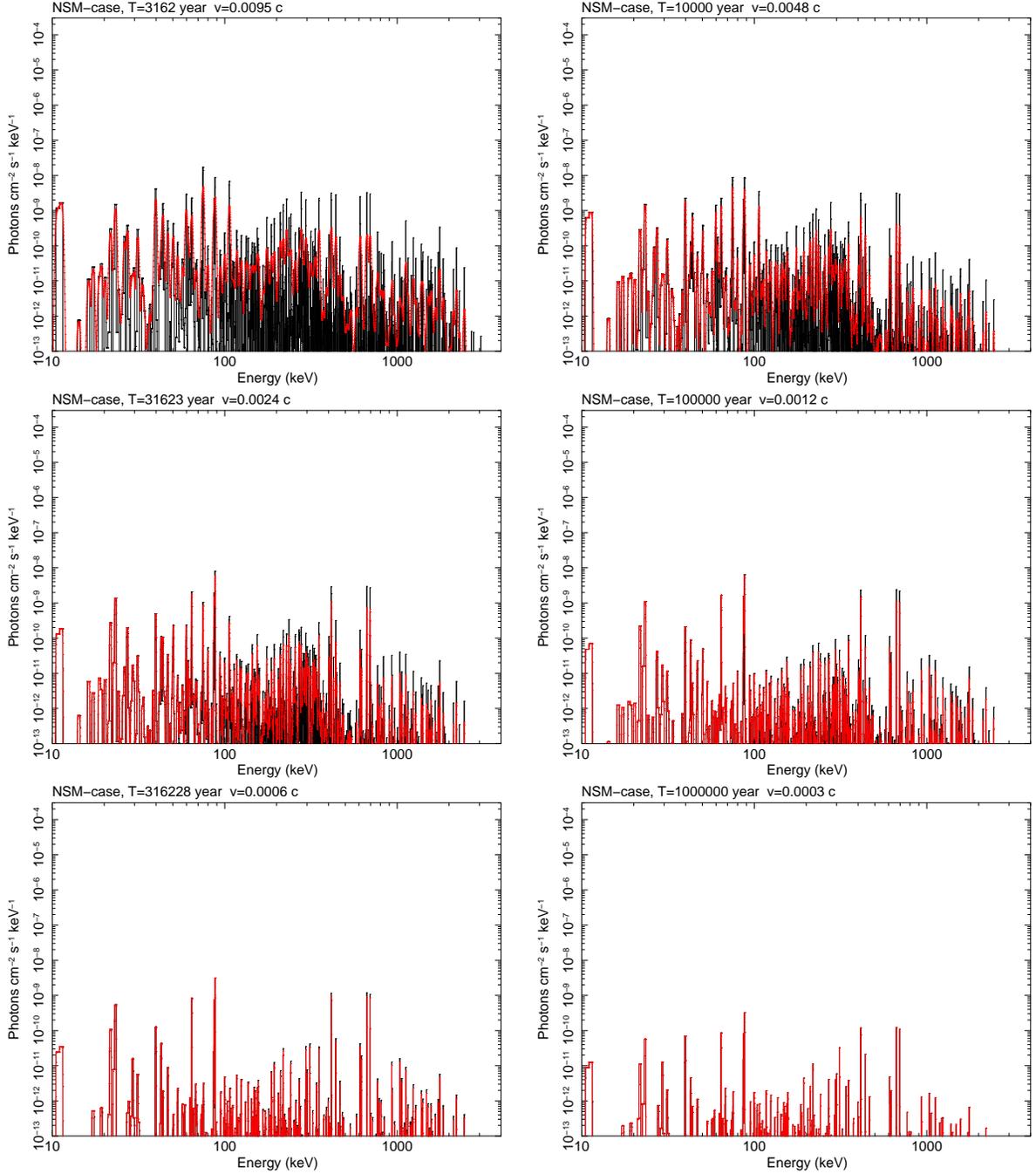

    \begin{center}
    \includegraphics[width=0.33 \textwidth, angle=-90]{gammasim_yterada_apj_figure05_3162yr.eps}
    \includegraphics[width=0.33 \textwidth, angle=-90]{gammasim_yterada_apj_figure05_10000yr.eps}
    \includegraphics[width=0.33 \textwidth, angle=-90]{gammasim_yterada_apj_figure05_31623yr.eps}
    \includegraphics[width=0.33 \textwidth, angle=-90]{gammasim_yterada_apj_figure05_100000yr.eps}
    \includegraphics[width=0.33 \textwidth, angle=-90]{gammasim_yterada_apj_figure05_316228yr.eps}
    \includegraphics[width=0.33 \textwidth, angle=-90]{gammasim_yterada_apj_figure05_1000000yr.eps}
    \end{center}
    \caption{Same as Figure \ref{fig:rprocess_g_model_plot1} but for $t=$ 3,162--1,000,000 years.}
    \label{fig:rprocess_g_model_plot2}
\end{figure*}

From the numerical calculation described in Section \ref{sec:estimation:summary},
the $\gamma$-ray spectra from $t= 3$ to 1 million years, 
under the assumption that the ejecta mass is $M_{\rm ej} = 0.01 M_\odot$ at $d = 10$ kpc,
are summarized in Figures \ref{fig:rprocess_g_model_plot1} and \ref{fig:rprocess_g_model_plot2}.
As described in Section \ref{sec:intro}, 
we assume that all of the emissions from the NSM remnants are observable within the wider FOV;
this is assumed to be larger than the object size,
which becomes around 10 pc at $t=1,000$ years and expands into around 100 pc at $t>10^6$ years.
The spectra contain many nuclear lines broadened by the Doppler effect. 
They appear to form a continuous spectrum in the early phase, but they become separated at ages older than 1,000 years.
Note that the $\gamma$-ray data without the Doppler-broadening effect 
(the outputs from the second step of the calculation in Section \ref{sec:estimation:summary})
is provided as the numerical model for the XSPEC tool \citep{1996ASPC..101...17A} in the HEAsoft package 
(Appendix \ref{sec:appendix:xspec}).

\begin{table}[ht]
    \centering
    \begin{tabular}{clll}
    \hline 
    Energy (keV) & Nuclei & Major $Y_{\rm e}$ & Half-life\\
    \hline 
    40.0    $\ddagger$ &  $^{225}$Ra   & 0.10--0.20  &  $\sim 7.9\times 10^3$ yr,\\
                       &               &             &  $1.6\times 10^5$ yr,\\
                       &               &             &  $2.1\times 10^8$ yr\\
     58.6   $\ddagger$ &  $^{60{\rm m}}$Co   & 0.35--0.40  &  $\sim 2.6\times 10^6$ yr \\
     59.5              &  $^{241}$Am   & 0.10--0.20  &  $\sim 4.3\times 10^2$ yr,\\
                       &               &             &  $8.4\times 10^3$ yr\\
     74.7   $\ddagger$ &  $^{243}$Am   & 0.15--0.20  &  $\sim 7.4\times 10^3$ yr\\
     87.6   $\dagger$  &  $^{126}$Sn   & 0.10--0.30  &  $\sim 2.3\times 10^5$ yr\\
    106.1   $\ddagger$ &  $^{239}$Np   & 0.10--0.20  &  $\sim 4.5\times 10^9$ yr\\
    236.0              &  $^{227}$Th   & 0.10        &  $\sim 21$ yr,\\
                       &               &             &  $3.2\times 10^4$ yr,\\
                       &               &             &  $7.0\times 10^8$ yr \\
    276.0              &  $^{81}$Kr    & 0.45        &  $\sim 2.2\times 10^5$ yr \\
    311.9              &  $^{233}$Pa   & 0.10--0.20  &  $\sim 2.1\times 10^6$ yr\\
    328.4              &  $^{194}$Ir   & 0.10--0.15  &  $\sim 6$ yr\\
    427.9   $\dagger$  &  $^{125}$Sb   & 0.25--0.30  &  $\sim 2.7$ yr\\
    440.5   $\ddagger$ &  $^{213}$Bi   & 0.10--0.20  &  $\sim 7.3\times 10^3$ yr\\
    511.9              &  $^{106}$Rh   & 0.30--0.35  &  $\sim 1.1$ yr\\
    513.9   $\dagger$  &  $^{85}$Kr    & 0.30--0.40  &  $\sim 10$ yr\\
    561.1, 834.5       &  $^{92}$Nb    & 0.45          &  $\sim 3.4\times 10^7$ yr \\
    609.3   $\ddagger$ &  $^{214}$Bi   & 0.10--0.20  &  $\sim 1.6\times 10^3$ yr,\\
                       &               &             &  $\sim 7.5\times 10^4$ yr\\
    661.7   $\dagger$  &  $^{137{\rm m}}$Ba  & 0.10--0.25  &  $\sim 30$ yr\\
    765.8              &  $^{95}$Nb    & 0.30--0.35  &  $\sim 34$ d, 64 d\\
    871.1, 702.6       &  $^{94}$Nb    & 0.45          &  $\sim 2.0\times 10^4$ yr \\
    1115.5             &  $^{65}$Zn    & 0.45          &  $\sim 243$ d\\
    1157.0             &  $^{44}$Sc    & 0.45          &  $\sim 60$ yr \\
    1332.5             &  $^{60}$Co    & 0.45          &  $\sim 5.2$ yr\\
    \hline 
    \end{tabular}
    \caption{List of nuclei and their most prominent nuclear $\gamma$-ray lines, arranged by $Y_{\rm e}$.
    The $\dagger$ and $\ddagger$ marks indicate the lines used in the $Y_{\rm e}$ diagnostics 
    in Section \ref{sec:diagnostic:ye} for young ($t<100$ years) and old ($t> 100$ years) cases, respectively. Note that the 'm' in $^{137{\rm m}}$Ba indicates that it is meta-stable.}
    \label{tab:bright_lines}
\end{table}

To identify the $\gamma$-lines in the spectra, 
we checked the most prominent lines in the $\gamma$-ray spectra 
generated by a single $Y_{\rm e}$ condition.
Table \ref{tab:bright_lines} lists the brightest lines shown for the nuclei in each $Y_{\rm e}$.
Roughly speaking, the bright lines seen for objects of a younger age lie in the higher-energy $\gamma$-ray band of the spectrum.
Further details of the diagnostics will be discussed in Section \ref{sec:diagnostic}.

\section{Gamma-ray Diagnostics of Neutron-Star Merger Remnants}
\label{sec:diagnostic}
\subsection{Spectral Color Changes of Neutron-Star Merger Remnants}
\label{sec:diagnostic:color}
\begin{figure}[ht]
    \centering
    \includegraphics[width=0.45 \textwidth]{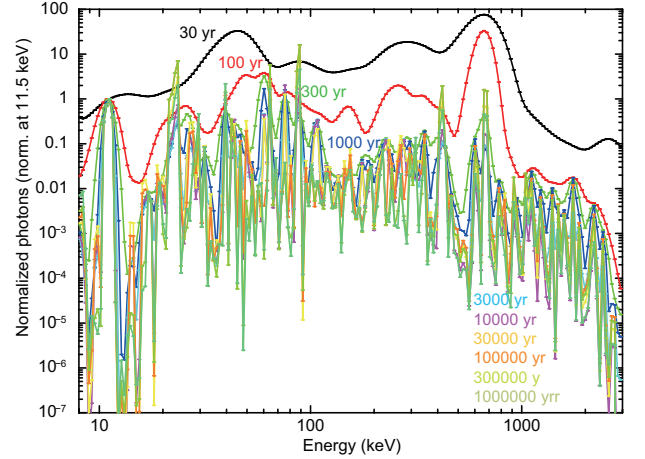}
    \caption{The energy spectra with Doppler broadening, same as the red plots in Figures \ref{fig:rprocess_g_model_plot1} and \ref{fig:rprocess_g_model_plot2} but normalized to the 11.5 keV flux at 30, 100, 300, 1000, 3000, 10000, 30000, 100000, 300000, and 1000000 years, which are shown in black, red, green, blue, light blue, magenta, yellow, orange, yellow green, and olive green, respectively.}
    \label{fig:normalized_spectra}
\end{figure}

Using the energy spectra of the NSM remnants
(Figures \ref{fig:rprocess_g_model_plot1} and \ref{fig:rprocess_g_model_plot2}), 
we first checked the properties of the spectral shapes 
from the hard X-ray to the soft $\gamma$-ray bands.
As shown in the normalized spectra plotted in Figure \ref{fig:normalized_spectra},
the energy spectra roughly evolve from hard to soft slopes. 
$\gamma$-ray emission decreases rapidly 
leaving the hard X-ray emission in old age,
as is indicated in Table\ref{tab:bright_lines}.
This phenomenon of $\gamma$ rays is equivalent to the Sargent law for $\beta$ decay.

\begin{figure}[hbt]
    \centering
    \includegraphics[width=0.45 \textwidth]{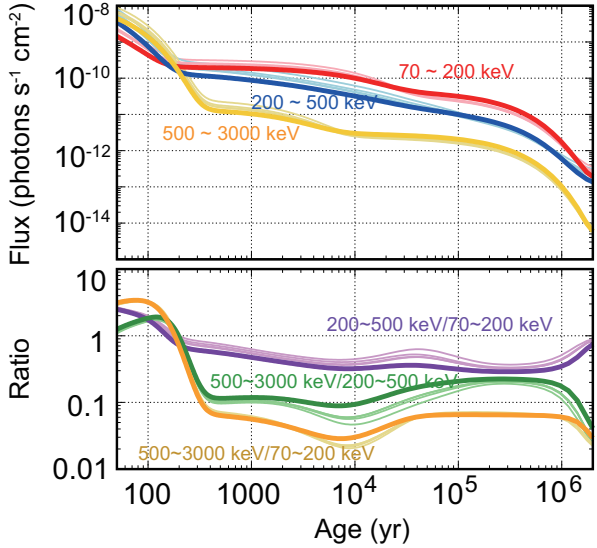}
    \caption{The top panel represents the light curve of the $\gamma$-ray spectra simulated for the case of an NSM at a distance of 10 kpc, with the Doppler-broadening effect and an initial velocity of 0.3 $c$. The energy bands of 70--200 keV, 200--500 keV, and 500--3000 keV, are shown in red, blue, and yellow, respectively. 
    The bottom panel represents the ratios among the three bands in the top panel: the ratio between the 70--200 keV and 200--500 keV bands, that between the 200--500 keV and 500--3000 keV bands, and that between the 70--200 keV and 500--3000 keV bands are shown in purple, green, and orange, respectively.
    The dim colors represent the results using the $Y_{\rm e}$-fraction by \citet{2014ApJ...789L..39W} as the nomical condition of this paper, and the lighter color lines represent the results by the other models, DD2-125145, DD2-135135, SFHo-125145, and SFHo-135135 in \citet{2022MNRAS.510.2804K}.}
    \label{fig:energy_resolved_lc_ratio}
\end{figure}

To see the evolution of the shape of the $\gamma$-ray spectra more quantitatively,
we plotted the light curves of the $\gamma$-ray flux in three bands:
70 to 200 keV, 200 to 500 keV, and 500 to 3,000 keV,
which cover multiple lines around 100 keV and 300 keV, and a prominent line around 700 keV,
respectively. 
As indicated in the top panel of Figure \ref{fig:energy_resolved_lc_ratio},
the flux in the higher-energy bands decreases more quickly than that in the low-energy bands.
A decaying trend is also seen in the time dependency of the hardness ratio among these bands,
as is indicated in the lower panel of Figure \ref{fig:energy_resolved_lc_ratio}.
The ratio drops dramatically at around 200--300 years,
indicating that the $\gamma$-ray flux above the 500 keV band quickly decreases at this age.
This phenomenon is primarily due to the decay of $^{125}$Sb and $^{137{\rm m}}$Ba 
listed in Table \ref{tab:bright_lines}.
Note that this result does not change even if we adopt other $Y_{\rm e}$-fraction models of DD2-125145, DD2-135135, SFHo-125145, and SFHo-135135 in \citet{2022MNRAS.510.2804K}, as shown in Figure \ref{fig:energy_resolved_lc_ratio}.

\begin{figure}[ht]
    \centering
    \includegraphics[width=0.45 \textwidth]{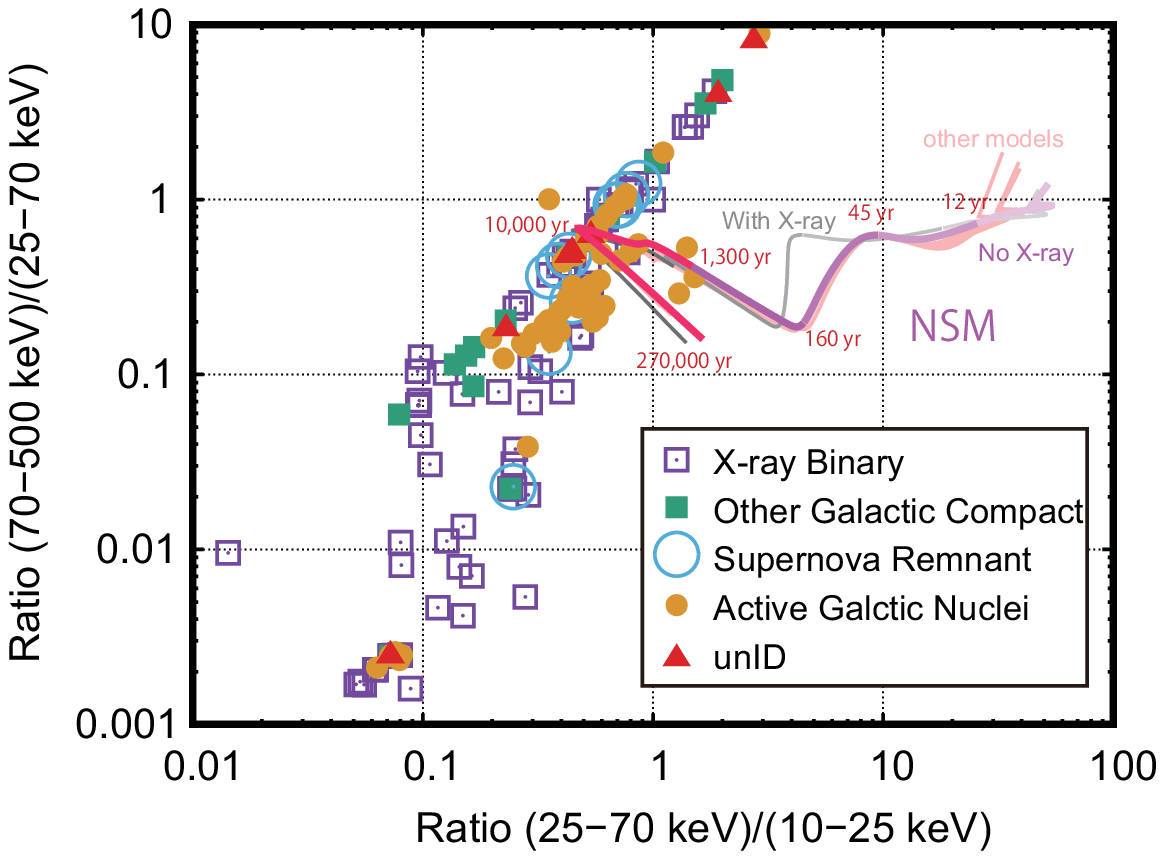}
    \includegraphics[width=0.43 \textwidth]{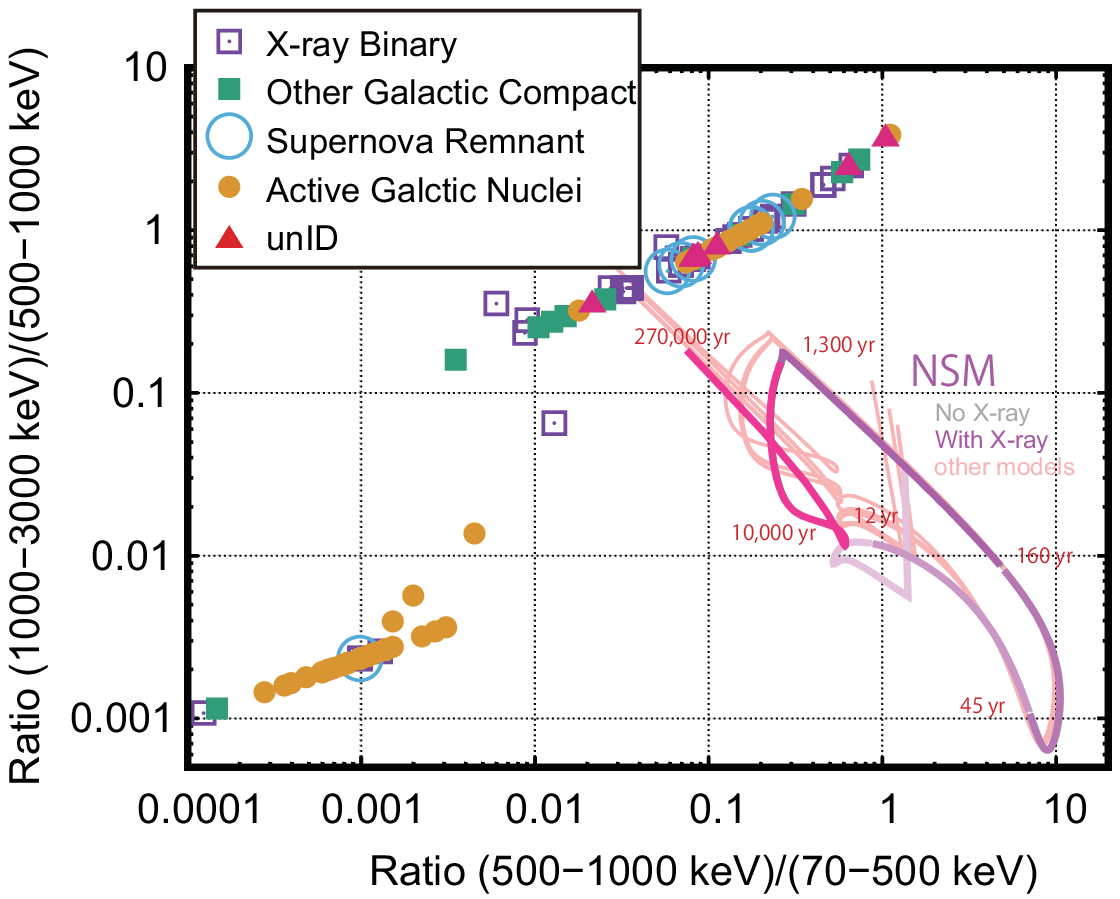}
    \caption{(Top) A color-color diagram of the flux ratio between the 10--25 keV and 25--70 keV bands versus that between the 25--70 keV and 70--500 keV bands is shown using simulated $\gamma$-ray spectra for the case of an NSM at a distance of 10 kpc, with the Doppler-broadening effect and an initial velocity of 0.3 $c$.
    The spectral evaluations of the NSM both with and without K X-ray emission are shown by magenta and gray lines, respectively, 
    and those by the other $Y_{\rm e}$-fraction models, DD2-125145, DD2-135135, SFHo-125145, and SFHo-135135 in \citet{2022MNRAS.510.2804K}, are shown in the light-pink lines, for reference.
    The color-color diagram of the X-ray objects listed in the INTEGRAL catalog (version 0043) are also plotted: X-ray binaries with the purple squares, galactic compact objects, such as cataclysmic variables, with the green squares, supernova remnants with the cyan circles, active galactic nuclei with the orange circles, and unidentified objects with the red triangles.
    (Bottom) Same plot as the top panel but in the 70--500, 500--1,000, and 1,000--3,000 keV bands.}
    \label{fig:spectral_color}
\end{figure}

To compare the spectral shape of NSM remnants with other astronomical objects,
we plotted the color-color diagrams 
in the hard X-ray band (10--500 keV) 
and in the hard X-ray to $\gamma$-ray band (70--3000 keV), 
in the top and bottom of Figure \ref{fig:spectral_color}, respectively.
We divided the energy band-pass for these spectra into three ranges: 
10--25, 25--70, and 70--500 keV 
for the hard X-ray band (top  of Figure \ref{fig:spectral_color}),
and 70--500, 500--1,000, and 1,000--3,000 keV 
for the hard X-ray to $\gamma$-ray band (Figure \ref{fig:spectral_color} bottom).
Note that the divisions of the energy bands are defined so that they follow the energy band-pass of current $\gamma$-ray instruments on board NuSTAR \citep{2013ApJ...770..103H}, INTEGRAL \citep{2003A&A...411L...1W}, and other observatories.
For comparison, the spectral colors of other astronomical objects, calculated using the INTEGRAL catalog version 0043
\footnote{https://www.isdc.unige.ch/integral/science/catalogue}, are also plotted in the same figures.
In the hard X-ray band (the 10--500 keV band in the top of Figure \ref{fig:spectral_color}), 
the spectra of NSM remnants older than $t\sim$ 1,000 years have spectral colors 
similar to those of supernova remnants or active galactic nuclei, 
but NSM remnants younger than $t\sim$ 1,000 years can be distinguished 
from other known objects by their hard X-ray colors.
In other words, the spectral color in the hard X-ray band below 500 keV
is a good indicator of young NSM remnants.
Furthermore, this differentiation from known objects becomes more prominent when we include the higher-energy band covering the MeV portion of the spectrum, as is clearly indicated in the bottom of Figure \ref{fig:spectral_color}.
Note that this result does not change even if we adopt other $Y_{\rm e}$-fraction models of DD2-125145, DD2-135135, SFHo-125145, and SFHo-135135 in \citet{2022MNRAS.510.2804K}, as shown in Figure \ref{fig:spectral_color}.
Therefore, NSM remnants have unique spectral colors in the hard X-ray to $\gamma$-ray bands.
This observation is one of the important conclusions from our calculation.
Note that the spectral models in the INTEGRAL catalog are simple enough that 
the colors of known objects in the gamma-ray band 
(bottom of Figure  \ref{fig:spectral_color}) are less scattered 
than those in the hard X-ray band (top of Figure \ref{fig:spectral_color}).
The spectral separation between NSM remnants and other objects in the bottom of Figure \ref{fig:spectral_color} does not change dramatically,
even if we lower the low-energy threshold (70 keV in the bottom of Figure \ref{fig:spectral_color}) to cover 20 keV, for example. 
However, it becomes worse if we set it higher so that everything up to a certain point, 200 keV, for example, is ignored.
This implies that hard X-rays around 100 keV provide the key information
for distinguishing NSM remnants from other objects. 
Note that these results are based on the pure-nuclear $\gamma$ rays from $r$-process nuclei in NSMs, and thus the synchrotron radiation from electrons that are accelerated by the shocks may contaminate the hard X-ray band for young remnants. 
Additionally, when taking actual observations, we must be careful to isolate the contamination of the hard X-ray spectrum that arises from other objects located behind the NSM, such as active galactic nuclei within the FOV.

\subsection{Nuclear Line Emissions from Older Neutron-Star Merger Remnants}
\label{sec:diagnostic:lines}

For ages older than $t>$ 3,000 years, 
nuclear lines are clearly seen in the $\gamma$-ray spectra of NSM remnants due to the minimal Doppler-broadening effect,
as is shown in Figures \ref{fig:rprocess_g_model_plot1} and \ref{fig:rprocess_g_model_plot2}.
Using the $\gamma$-ray spectra of NSM remnants that were shown in Section \ref{sec:estimation}
(i.e., the $Y_{\rm e}$ distribution for the NSM case with $M_{\rm ej} = 0.01 M_\odot$ at $d = 10$ kpc),
we selected the brightest nuclear lines in each energy band, 3--75 keV, 75--500 keV, and 500--4000 keV, for at least one epoch in the age range spanning $t=$10 to $4 \times 10^6$ years.
Note that these energy bands are defined in such that they simulate the energy bands that are observable by current and near-future instruments aboard satellites, such as the hard X-ray focusing missions NuSTAR \citep{2013ApJ...770..103H} and FORCE \citep{2018SPIE10699E..2DN}, and $\gamma$-ray missions like INTEGRAL \citep{2003A&A...411L...1W}, e-ASTROGAM \citep{2017ExA....44...25D,2018JHEAp..19....1D}, AMEGO \citep{2020SPIE11444E..31K}, and GRAMS \citep{2020APh...114..107A}.

\begin{figure}[hbt]
    \centering
    \includegraphics[width=0.45 \textwidth]{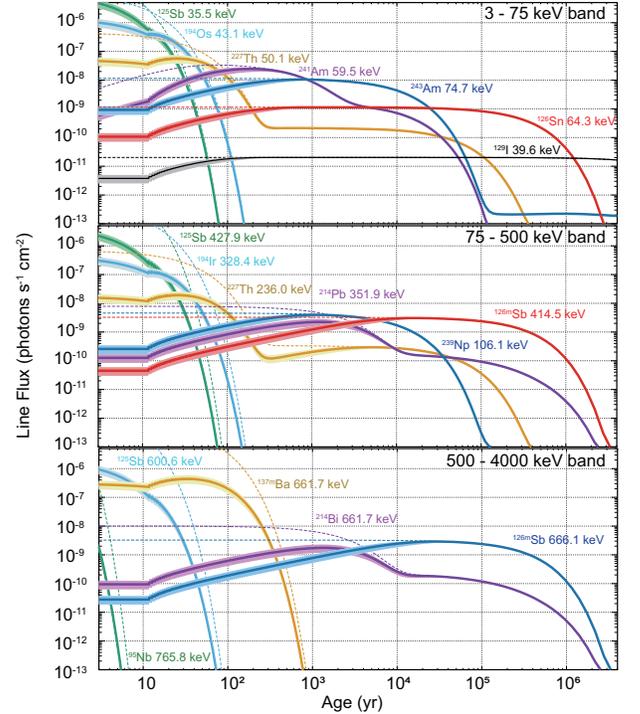}
    \caption{The time evolution of the brightest $\gamma$-ray lines
    in the energy bands of 3--75 keV, 75--500 keV, and 500--4000 keV
    are shown in the top, middle, and bottom panels, respectively.
    The line fluxes with and without the Doppler effect are shown 
 as thick and dashed lines, respectively.
    The thick lines plot the $\gamma$-ray fluxes within the $\Delta E = 3$ keV width, 
    and the upper and lower boundaries of the corresponding hatched areas 
show the flux within the $\Delta E = 2$ keV and 4 keV widths, respectively. 
    The nuclei and energies are indicated by the same color as the lines and hatches.}
    \label{fig:line_history_3bands}
\end{figure}

Figure \ref{fig:line_history_3bands} presents the time evolution 
of the brightest nuclear $\gamma$-ray lines in these energy bands.
To account for the reduction in the line sensitivities as a result of the Doppler-broadening effect,
we accumulated the photons that were within the energy resolution of $\Delta E = 3 \pm 1$ keV from the center energy of their associated lines.
This chosen value for the energy resolution is typical for semiconductor $\gamma$-ray detectors.
For reference, the evolution of lines without Doppler broadening is also shown in the figure as dashed lines.
As indicated in Figure \ref{fig:line_history_3bands},
the Doppler-broadening effect becomes less dominant in the hard X-ray band
after a few hundred years,
but it is still present until about $t=10^3$ and $10^4$ years
in the soft $\gamma$-ray and the hard $\gamma$-ray bands, respectively.
Note that the reason why several lines, such as those of $^{126{\rm m}}$Sb and $^{239}$Np, increase as $t$ approaches $10^3$--$10^5$ years is that the number of parent nuclei increases in these phases.

From Figure \ref{fig:line_history_3bands}, 
we can identify the nuclear lines that are useful as indicators for the ages of NSMs.
The ages can be categorized into three epochs: 
$t < 100$ years, $t \sim 10^{3}$-- $10^4$ years, and $t>10^4$ years.
In summary, if we detect the lines from $^{125}$Sb, $^{194}$Os, $^{227}$Th, or $^{194}$Ir, 
then we can determine the age of the NSM to be very young at $t < 100$ years.
Similarly, lines from $^{137{\rm m}}$Ba in the $\gamma$-ray band indicate that the age is around 
$t \sim 10^{2}$ years.
In the age range spanning $t \sim 10^{3}$-- $10^4$ years, 
nuclear lines will be detected from $^{241}$Am, $^{243}$Am, $^{214}$Pb, $^{239}$Np, and/or $^{214}$Bi.
A nuclear line from $^{126{\rm m}}$Sb indicates that 
the NSM is very old at $t>10^4$ years.
In the wide age range from $t=400$ to $10^5$ years, the line from $^{126}$Sn stays almost constant at $10^{-9}$ photons s$^{-1}$ cm$^{-2}$ for a distance of $d=10$ kpc, and thus it can be used as a standard candle for measuring $d$.

\subsection{Line Diagnostics for the electron fraction}
\label{sec:diagnostic:ye}

In addition to the spectral colors (Section \ref{sec:diagnostic:color}),
nuclear lines can be used to identify NSM remnants among astronomical objects,
especially when the remnants are of an older age.
Since the NSMs are thought to have both a more neutron-rich environment 
and a lower $Y_{\rm e}$ condition than SNs
\citep{2011ApJ...726L..15W,1974ApJ...192L.145L,2010MNRAS.406.2650M},
a new line-diagnostic method utilizing $Y_{\rm e}$ values will 
be useful for distinguishing NSMs from SNe. 
In this subsection, we search for $\gamma$-ray line diagnostics for $Y_{\rm e}$.
We use the $\gamma$-ray spectra calculated under the pure $Y_{\rm e}$ conditions in the $Y_{\rm e} = 0.10$--0.45 range,
whereas in the previous sections we used the mixed $Y_{\rm e}$ condition for NSMs.

To identify the best candidates among the nuclear $\gamma$-ray lines for the identification of $Y_{\rm e}$,
we first selected the five brightest lines for each age, $t=100, 1000, 10^4, 10^5$, and $10^6$ years, 
and for each $Y_{\rm e}$ ($= 0.10, 0.15, 0.20, 0.25, 0.30, 0.35, 0.40$, and $0.45$).
Then among these $5$ (ranks) $\times 5$ ($t$) $\times 7$ ($Y_{\rm e}$) lines, 
we selected the nuclear lines which appeared in two or more of the conditions for $t$ and $Y_{\rm e}$.
In total, ten $\gamma$-ray lines are selected
and are marked as $\dagger$ and $\ddagger$ in Table \ref{tab:bright_lines} 
for $t <$ 100 years and $t >$ 100 years, respectively. 
Therefore, the lines from $^{137{\rm m}}$Ba (661.66 keV), $^{85}$Kr (513.9 keV), and $^{125}$Sb (427.87 keV) 
in ages below 100 years are indicators 
of low, middle, and high $Y_{\rm e}$ environments, respectively.
Here, low, middle, and high are numerically defined as 
$Y_{\rm e} \sim 0.10$--$0.20$, $0.20$--$0.35$, and $0.35$--$0.45$, respectively.
In ages older than $t=$100 years,
the lines from $^{225}$Ra (40.0 keV), $^{243}$Am (74.66 keV), 
$^{239}$Np (106.1 keV), $^{213}$Bi (440.5 keV), and $^{214}$Bi (609.3 keV) 
are emitted from a low $Y_{\rm e}$ environment,
whereas the lines from $^{60{\rm m}}$Co (58.6 keV) and $^{126}$Sn (87.6 keV)
become bright in the middle and high $Y_{\rm e}$ environments, respectively.

\begin{figure*}[hb]
    \centering
    \includegraphics[width=0.55 \textwidth]{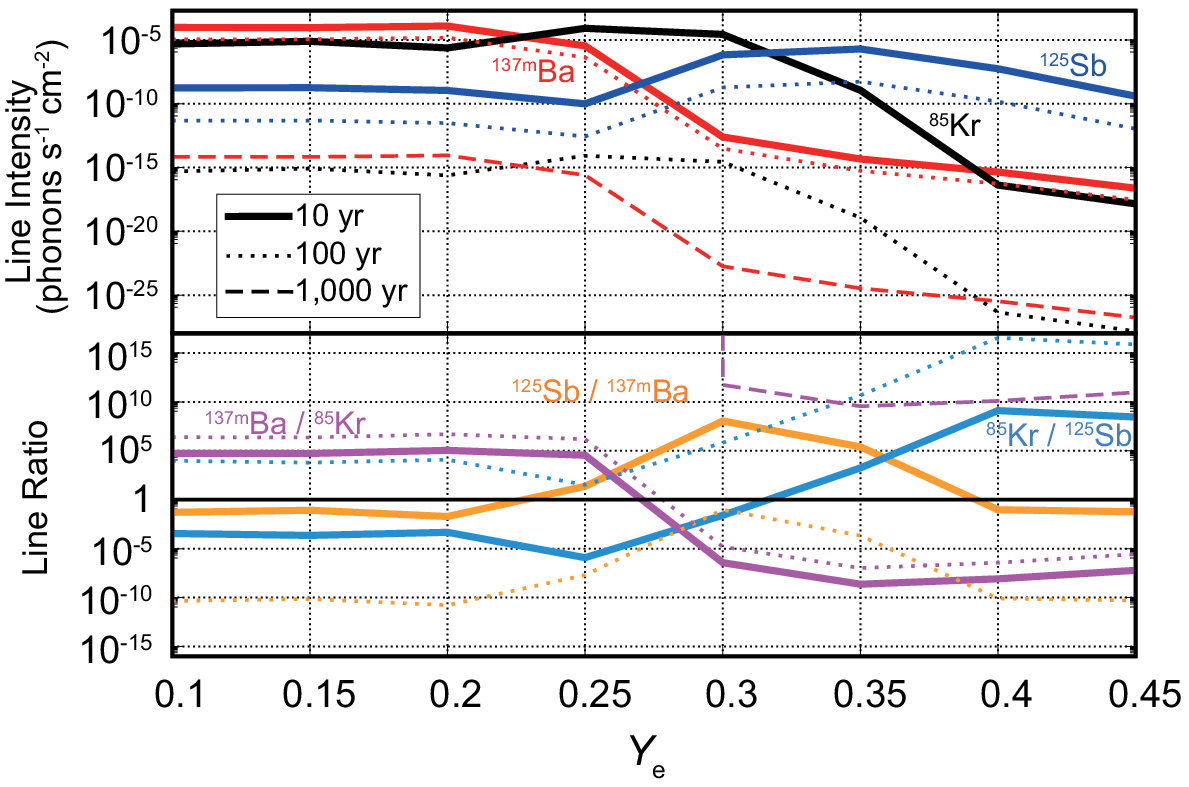}\\
    \includegraphics[width=0.55 \textwidth]{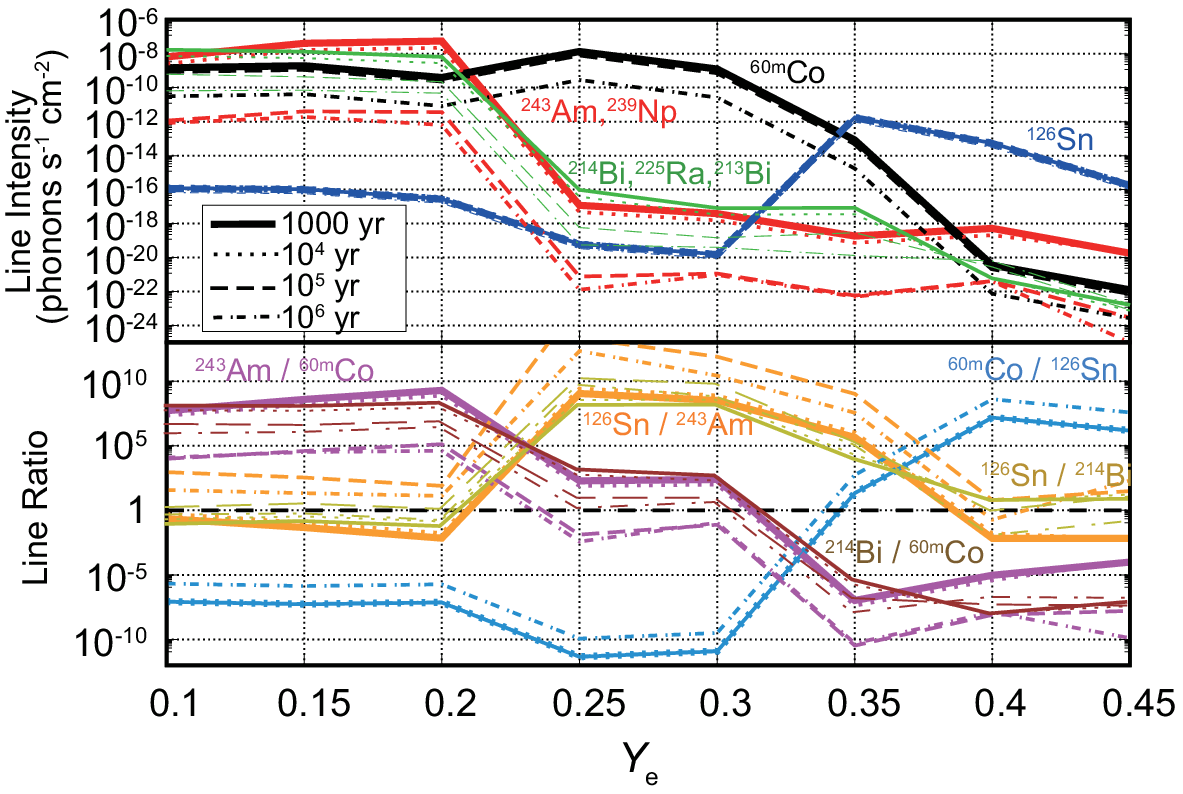}\\
    \includegraphics[width=0.52 \textwidth]{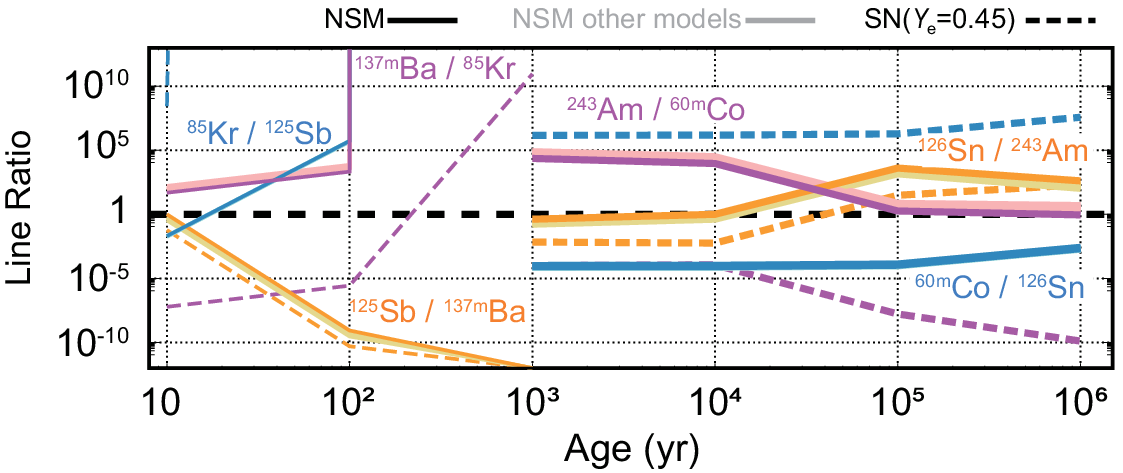}
   \caption{(Top) The $Y_{\rm e}$ dependencies of the line intensities of the nuclei listed 
    with $\dagger$ marks in Table \ref{tab:bright_lines} are shown in the upper panel.
    The reduction due to the Doppler-broadening effect is not considered.
    The line intensities of $^{137{\rm m}}$Ba, $^{85}$Kr, and $^{125}$Sb are shown in red, black, and blue, respectively. 
    Thick, dotted, and dashed lines represent the intensities at $t =$ 10, 100, and 1000 years. 
    The ratios among the lines are plotted in the lower panel; 
    the ratios between $^{137{\rm m}}$Ba and $^{85}$Kr, $^{125}$Sb and $^{137{\rm m}}$Ba, and $^{85}$Kr and $^{125}$Sb are shown in purple, orange, and cyan, respectively.
    (Middle) Same plot as the top panel but for older ages ($t>100$ years, $\ddagger$ marks in Table \ref{tab:bright_lines}).
    The dependencies of $^{243}$Am (and $^{239}$Np), $^{214}$Bi (and $^{225}$Ra, $^{213}$Bi), $^{60{\rm m}}$Co, and $^{126}$Sn are shown in red, green, black, and blue, respectively, and 
    the thick, dotted, dashed, and dotted dash lines represent the intensity at $t =$ 1000, $10^4$, $10^5$, and $10^6$ years, respectively, in the upper panel.
    The ratios in the lower panel between $^{243}$Am and $^{60{\rm m}}$Co, between $^{214}$Bi and $^{60{\rm m}}$Co, between $^{126}$Sn and $^{243}$Am, between $^{126}$Sn and $^{214}$Bi, and between $^{60{\rm m}}$Co and $^{126}$Sn are shown in purple, brown, orange, dark-yellow, and cyan, respectively.
    (Bottom) The time dependencies of the line ratios of $^{137{\rm m}}$Ba/$^{85}$Kr, $^{125}$Sb/$^{137{\rm m}}$Ba, $^{85}$Kr/$^{125}$Sb, $^{243}$Am/$^{60{\rm m}}$Co, $^{126}$Sn/$^{243}$Am, and $^{60{\rm m}}$Co/$^{126}$Sn are shown in the thin purple, thin orange, thin cyan, thick purple, thick orange, and thick cyan lines, respectively. The NSM case by the mass fraction of \citet{2014ApJ...789L..39W} and the case of $Y_{\rm e}=0.45$ are shown in the straight and dotted lines, respectively. The results by the other NSM models, DD2-125145, DD2-135135, SFHo-125145, and SFHo-135135 in \citet{2022MNRAS.510.2804K}, are shown in the lighter colors.}
    \label{fig:ye_dependency}
\end{figure*}

Since the absolute flux of a single line changes with respect to $t$ and $d$,
the ratio between two or more lines should be a good indicator for $Y_{\rm e}$.
Figure \ref{fig:ye_dependency} summarizes the line intensities and their ratios using the ten nuclei selected above.
For simplicity, the plots for young and old ages, $t=$ 10--$10^3$ years and $10^3$--$10^6$ years, respectively, are shown separately.
In the young age range (top of Figure \ref{fig:ye_dependency}), 
the ratios of $^{137{\rm m}}$Ba/$^{85}$Kr, $^{125}$Sb/$^{137{\rm m}}$Ba, and $^{85}$Kr/$^{125}$Sb
become larger than unity in the low, middle, and high $Y_{\rm e}$ environments, respectively.
These low and high $Y_{\rm e}$ indicators (i.e., $^{137{\rm m}}$Ba/$^{85}$Kr and $^{85}$Kr/$^{125}$Sb, respectively) 
exhibit more prominent ratios over time since $^{85}$Kr decays slower than $^{125}$Sb and faster than $^{137{\rm m}}$Ba,
whereas the middle indicator ($^{125}$Sb/$^{137{\rm m}}$Ba) becomes dim after 100 years.
Note that the plots use the incident line fluxes calculated in step 2 of Section \ref{sec:estimation:summary}, and the reduction due to the Doppler-broadening effect is not considered. 
The Doppler effect is particularly significant in plots for the young age range (top of Figure \ref{fig:ye_dependency}). Quantitatively, the ratios of 
$^{137{\rm m}}$Ba/$^{85}$Kr, $^{125}$Sb/$^{137{\rm m}}$Ba, and $^{85}$Kr/$^{125}$Sb
change by factors of $1.57, 0.82$, and $0.77$, respectively, for $t<1,000$ years.
In the old age range (middle of Figure \ref{fig:ye_dependency}),
the ratios of $^{243}$Am/$^{60{\rm m}}$Co, $^{126}$Sn/$^{243}$Am, and $^{60{\rm m}}$Co/$^{126}$Sn
indicate the low, middle, and high $Y_{\rm e}$ environments, respectively.
The line from $^{239}$Np has the same flux and time evolution as that from $^{243}$Am (red plots),
because they are in the same decay chain.
Similarly, the lines from $^{214}$Bi, $^{225}$Ra, and $^{213}$Bi (green plots) follow almost the same trend
as those from $^{243}$Am and $^{239}$Np (red plots).
Among them, the low $Y_{\rm e}$ indicator ($^{243}$Am/$^{60{\rm m}}$Co) in the old age range 
is valid up to $t=1$ million years, and 
the middle $Y_{\rm e}$ indicator ($^{126}$Sn/$^{243}$Am) shows more significant ratios with $t>10^3$ years.
On the other hand, lines for the high $Y_{\rm e}$ indicator $^{60{\rm m}}$Co/$^{126}$Sn 
decay quickly and become unavailable after $10^3$ years; 
that is, if the ratio $^{60{\rm m}}$Co/$^{126}$Sn is larger than unity,
then the object is in a high $Y_{\rm e}$ environment with an age of $t\sim 10^3$ years.
In summary, using these indicators, which become larger than unity in specific $Y_{\rm e}$ conditions, 
we can estimate the $Y_{\rm e}$ environment independently from the spectral-color diagnostics shown in Section \ref{sec:diagnostic:color}.

Finally, we checked the line ratios blended by $Y_{\rm e}$ distributions of the NSM cases.
The time evolution is plotted in Figure \ref{fig:ye_dependency} bottom. 
The difference of the $Y_{\rm e}$-fraction models between \citet{2014ApJ...789L..39W} and \citet{2022MNRAS.510.2804K} does not affect the trend of the NSMs so much.
These $\gamma$-ray lines are also expected to be observed from the remnants of core-collapse SNe,
which are considered to be less neutron-rich environment at $Y_{\rm e} \sim 0.5$ \citep{ 2020ApJ...890...35A} than NSMs.
However, our numerical-calculation model in this paper has limitations in estimating $\gamma$-ray radiation from the core-collapse SNe, because the mass fractions of the $r$-process nuclei in the ejecta are different between the maximum $Y_{\rm e}$ condition of our calculation (i.e., $Y_{\rm e} = 0.45$) and the SNe case ($Y_{\rm e} \sim 0.50$), and the neutron-rich nuclei in the nominal core-collapse SNe are predominantly generated via the $s$-process rather than the $r$-process.
For reference, we plotted the time evolution of the line ratios in the $\gamma$-ray spectra 
of $Y_{\rm e} = 0.45$, which should still reproduce well an environment with almost-equal numbers of neutrons and protons.
According to Figure \ref{fig:ye_dependency} bottom, we expect the low-$Y_{\rm e}$ indicators (i.e., $^{137{\rm m}}$Ba/$^{85}$Kr and $^{243}$Am/$^{60{\rm m}}$Co) in the NSM case become many orders-of-magnitude larger than those in the SNe case. 
In the core-collapse SNe where neutron-rich nuclei are generated via the $s$-process, relatively large amount of $^{85}$Kr and almost no $^{243}$Am are synthesized. 
Therefore, the difference of these low-$Y_{\rm e}$ indicators between the NSMs and SNe cases are expected to become larger than Figure \ref{fig:ye_dependency} bottom.
As for the middle-$Y_{\rm e}$ indicators ($^{125}$Sb/$^{137{\rm m}}$Ba and $^{126}$Sn/$^{243}$Am), they may not be useful to distinguish $\gamma$-rays from NSMs and SNe according to Figure \ref{fig:ye_dependency} bottom.
In the $s$-process environment of core-collapse SNe, almost no $^{137{\rm m}}$Ba and $^{243}$Am are synthesized and thus theose middle-$Y_{\rm e}$ indicators can be larger than the values in the figure.
Finally, the high-$Y_{\rm e}$ indicators ($^{85}$Kr/$^{125}$Sb and $^{60{\rm m}}$Co/$^{126}$Sn) can discard the NSMs from the SNe cases as indicated by Figure \ref{fig:ye_dependency} bottom.
In summary, the new line-diagnostic method for $Y_{\rm e}$ provides a tool for distinguishing between NSMs and SNe.

\section{Discussion}
\label{sec:discussion}

In Section \ref{sec:estimation}, 
we presented a nuclear-decay simulation using a large nuclear database,
the goal of which was to estimate the $\gamma$-ray spectra of NSMs up to the age of $t=10^6$ years.
We have identified many nuclear lines, listed in Table \ref{tab:bright_lines},
that can be used for identifying the nucleosynthesis environments of NSMs,
even with the Doppler-broadening effect altering the profiles of these lines in the young age range.
In Section \ref{sec:diagnostic},
we numerically analyzed the simulated $\gamma$-ray spectra from NSMs 
and found that the spectral slope in the soft $\gamma$-ray band above 500 keV
changes at around $t=$ 200--300 years.
We also found that the spectral colors of NSMs in the hard X-ray to soft $\gamma$-ray bands
differ from those of other astronomical objects up to $t=10^5$ years old. 
Consequently, we can identify a $\gamma$-ray object as an NSM remnant
using the $\gamma$-ray spectral colors (Section \ref{sec:diagnostic:color}).
Among the many nuclear lines in the spectra,
we identified that
the nuclear lines from $^{241}$Am, $^{243}$Am, $^{214}$Pb, $^{239}$Np, and $^{214}$Bi
are prominent for $t=10^3$--$10^4$ years, and 
that the lines from $^{126}$Sn and $^{126{\rm m}}$Sb are prominent for $t>10^4$ years (Section \ref{sec:diagnostic:lines}).
In addition, we proposed a new line-diagnostic method for distinguishing $Y_{\rm e}$ environments
that uses the line ratios of $^{137{\rm m}}$Ba/$^{85}$K and $^{243}$Am/$^{60{\rm m}}$Co,
which become larger than unity for low $Y_{\rm e}$ objects with young and old ages, respectively (Section \ref{sec:diagnostic:ye}).
This diagnostic method distinguishes NSMs from SNe.  
In the next section, we focus on the sensitivities in the $\gamma$-ray band that are required
for current and future MeV $\gamma$-ray missions that aim  to search for Galactic NSM remnants.

\subsection{Detectable distance to Galactic Neutron-Star Merger Remnants}
\label{sec:discussion:detection}

\begin{figure}[hbt]
    \centering
    \includegraphics[width=0.45 \textwidth]{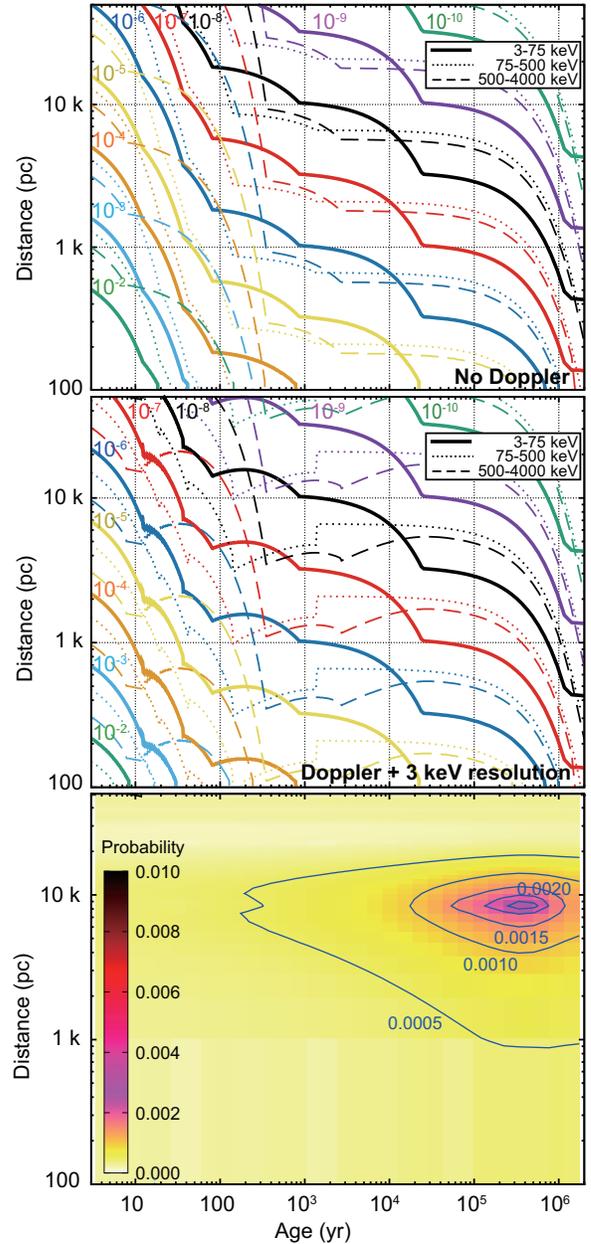}
    \caption{(Top, Middle) The distance limit of NSMs as a function of $t$ are shown 
    with the line sensitivities given in units of photons s$^{-1}$ cm$^{-2}$; see keys for details.
    The thick, dotted, and dashed lines represent the results 
    for the 3--75 keV, 75--500 keV, and 500--4,000 keV bands, respectively.
    The top and middle panels show the results without and with
    the Doppler-broadening effect, respectively.
    The line photons are accumulated within $\Delta E = 3$ keV in the middle panel.
    Note that the small jump in the 75--500 keV data at $t \sim 10^3$ years 
    in the middle panel is due to the interaction between the two brightest lines 
    when the Doppler-broadening effect is applied.
    (Bottom) Probabilities for the existence of NSMs, 
    which we took as an assumption in the calculation for Figure \ref{fig:plot_sensitivity},
    in the $t$--$d$ plane.}
    \label{fig:plot_distance}
\end{figure}

A $\gamma$-ray flux from the brightest line in a particular age range can be used to estimate the distance
that a virtual $\gamma$-ray instrument with a certain line sensitivity will be able to detect.
If we assume the age of the NSM remnants in Figure \ref{fig:line_history_3bands},
then we can estimate the limit of the distance that can be detected with the line sensitivity of a specific instrument.
The top of Figure \ref{fig:plot_distance} summarizes the achievable limit of $d$ for NSM remnants 
as a function of $t$ and is calculated for the three energy bands 
3--75 keV, 75--500 keV, and 500--4,000 keV.
For example, instruments with a line sensitivity of $10^{-7}$ photons s$^{-1}$ cm$^{-2}$ 
in the 3--75 keV band (red line in the top of Figure \ref{fig:plot_distance}), such as Hitomi HXI \citep{2014SPIE.9144E..25T} and NuSTAR \citep{2013ApJ...770..103H}, can observe the brightest lines from NSM remnants with $t<10^3$ years at $d=$ 3 kpc.
We also checked the degradation of the distance limit due to the Doppler-broadening effect, 
as shown in the middle of Figure \ref{fig:plot_distance},
but the results do not dramatically change.
If the line sensitivities are the same among the three energy bands,
then hard X-rays (thick lines) will be a powerful tool in the search for NSMs that are younger than $t<10^3$ years,
but the $\gamma$-ray observations (dotted or dashed lines) are better for surveying  NSMs that are older than $t>10^4$ years.

The G4.8+6.2 associated with AD~1163 is one example, from the middle of Figure \ref{fig:plot_distance},
that provides the requirement for the $\gamma$-ray sensitivity needed to observe an NSM remnant with a known distance and age.
The object is reported to be a young kilonova remnant with $t\sim 860$ years \citep{2019MNRAS.490L..21L}.
If it is an NSM remnant at $d\sim 10$ kpc, 
then a sensitivity of $10^{-8}$ and $10^{-9}$ photons s$^{-1}$ cm$^{-2}$ is required to observe G4.8+6.2 
in the hard X-ray and $\gamma$-ray bands, respectively.
This sensitivity is roughly one or two (or more) orders of magnitude deeper than that of INTEGRAL IBIS \citep{2003A&A...411L...1W}.
If the distance is closer at $d\sim 3$ kpc, 
then hard X-ray instruments with a sensitivity of $10^{-7}$ photons s$^{-1}$ cm$^{-2}$ in the 3--75 keV band, such as Hitomi HXI \citep{2014SPIE.9144E..25T} and NuSTAR \citep{2013ApJ...770..103H}, are expected to be able to observe the emissions from the object.

\subsection{Direct estimation of local NSM rates using gamma rays}
\label{sec:discussion:nsm_rate}

To estimate the coverage of Galactic NSM remnants observable for specific $\gamma$-ray sensitivities,
we first prepare a probability map for the existence of Galactic NSMs. 
This is given in the same plane as the top and middle of Figure \ref{fig:plot_distance} (the $t$--$d$ plane).
Since NSMs are not uniformly distributed in our Galaxy,
we apply the probabilities for NSMs in the $d$ and $t$ spaces given by \citet{2019ApJ...880...23W}
and multiply them to get the plot shown in the bottom of Figure \ref{fig:plot_distance}.
We assume that the NSMs are primarily concentrated around the Galactic plane.
Most of the NSMs are expected to exist at around $d \sim 8$ kpc
and $t \sim 10^4$--$10^6$ years,
as has already been described in \citet{2019ApJ...880...23W}.
 
\begin{figure}[hb]
    \centering
    \includegraphics[width=0.45 \textwidth]{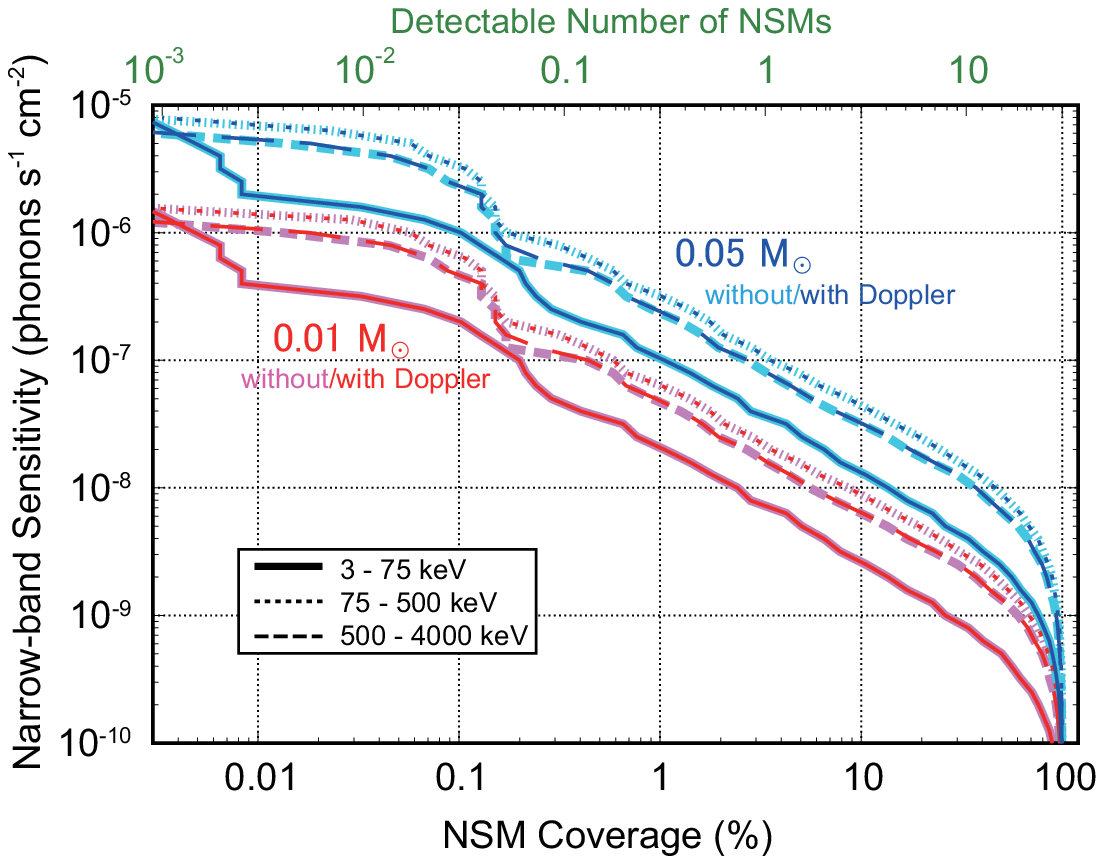}
    \includegraphics[width=0.45 \textwidth]{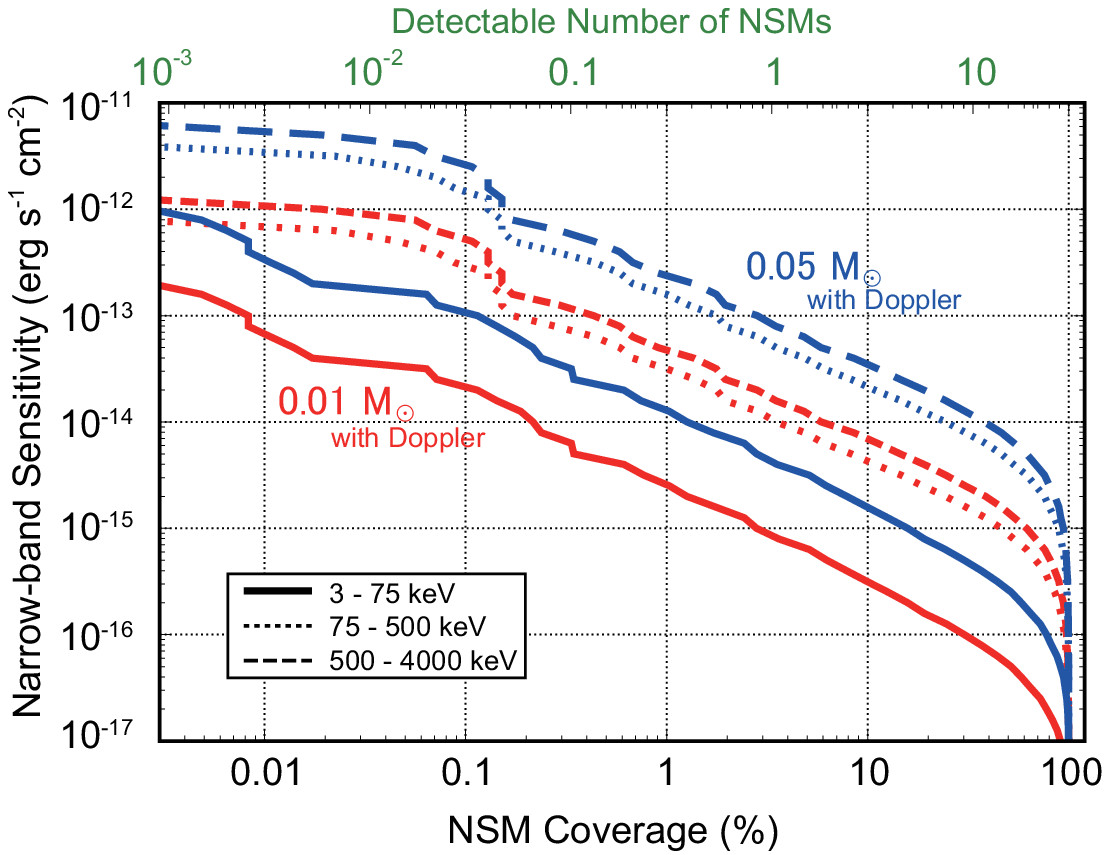}
    \caption{Coverage of NSM remnants in our Galaxy 
    as a function of the narrow-band sensitivity is shown.
    In the top panel, the results are given in units of photons s$^{-1}$ cm$^{-2}$.
    The magenta and red plots assume $M_{\rm ej} = 0.01 M_\odot$ with and without the Doppler effect, respectively.
    Similarly, the cyan and blue plots assume $M_{\rm ej} = 0.05 M_\odot$ with and without the Doppler effect.
    The thick, dotted, and dashed lines represent the coverage values
    in the 3--75 keV, 75--500 keV, and 500--4,000 keV bands, respectively.
    The top axis (shown in green) represents the detectable number of NSMs
    under the assumption of an NSM rate in our Galaxy of 30 per million years.
    The bottom panel shows the sensitivity results given in units of erg s$^{-1}$ cm$^{-2}$,
    but omitting the results without the Doppler effect.}
    \label{fig:plot_sensitivity}
\end{figure}

We then accumulate the probabilities for the existence of NSMs 
(bottom of Figure \ref{fig:plot_distance})
within the distance-limit curves (top and middle of Figure \ref{fig:plot_distance}).
As a result, we obtained the coverage of Galactic NSMs as a function of the line sensitivity;
this is shown in the top of Figure \ref{fig:plot_sensitivity}.
For example, if we survey Galactic NSM remnants with an instrument 
having a line sensitivity of $10^{-8}$ photons s$^{-1}$ cm$^{-2}$ in the 3--75 keV band,
then we expect to observe about 3\% of the NSMs in our Galaxy with $M_{\rm ej} = 0.01 M_\odot$.
This value corresponds to about one object if we assume an NSM rate in our Galaxy of 30 per million years.
In addition, we performed the same procedure to estimate the NSM coverage in units of erg s$^{-1}$ cm$^{-2}$,
which requires a sensitivity that is $E_{\rm \gamma}^1$ times higher than that required for units of photons s$^{-1}$ cm$^{-2}$.
$E_{\rm \gamma}$ here is the photon energy (the energy of the $\gamma$-ray line). 
The results are shown in the bottom of Figure \ref{fig:plot_sensitivity}.
Therefore, instruments that can achieve a sensitivity of $10^{-14}$ erg s$^{-1}$ cm$^{-2}$ in the 75--500 keV or the 500--4000 keV bands are expected to be able to observe one NSM remnant with $M_{\rm ej} = 0.01 M_\odot$ in our Galaxy under the same assumption of the NSM rate mentioned above.
Similarly, a sensitivity of $10^{-15}$ erg s$^{-1}$ cm$^{-2}$ is required in the hard X-ray band to observe one object with the same $M_{\rm ej}$.

\begin{figure}[ht]
    \centering
   \includegraphics[width=0.45 \textwidth]{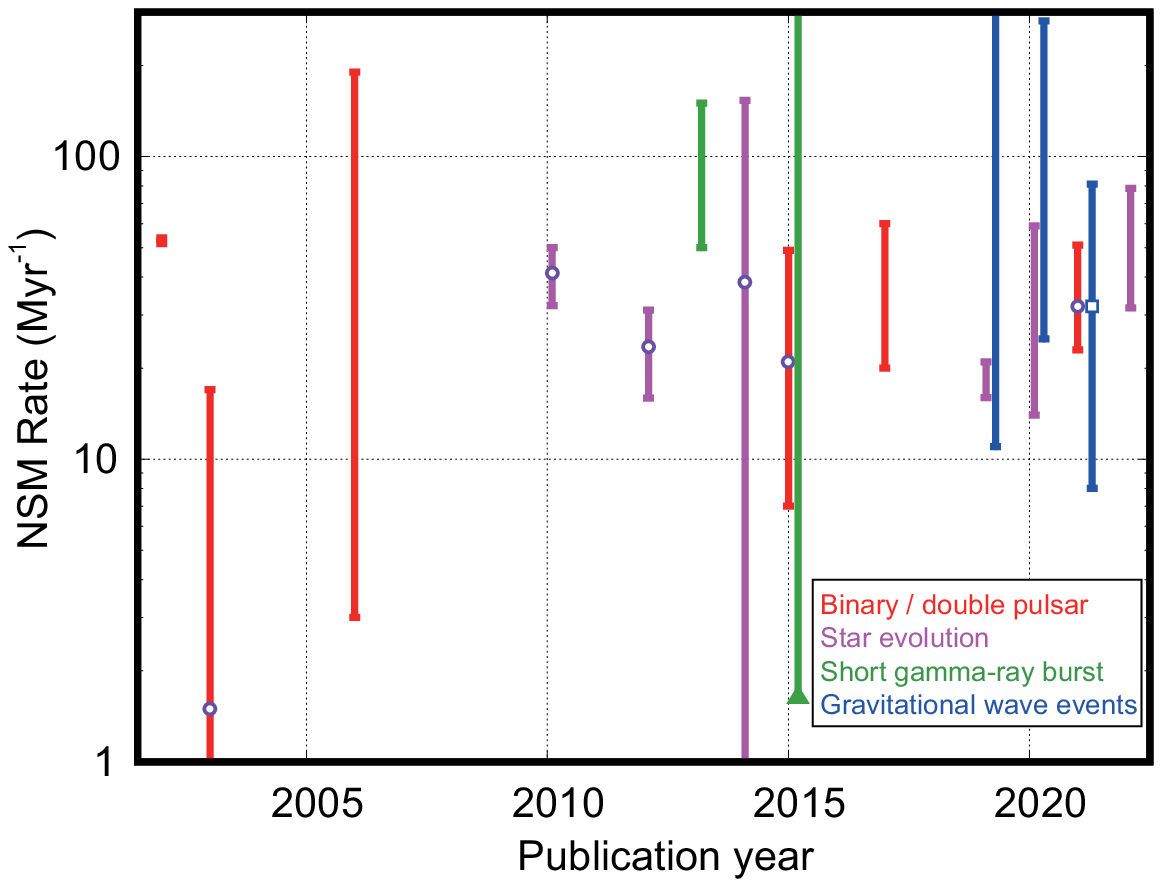}
    \caption{The NSM rate in our Galaxy estimated by previous studies as a function of the published year. 
    The red, magenta, green, and blue plots represent the estimation from 
    the binary or double pulsar population \citep{2002ApJ...572..407B, 2003MNRAS.342.1169V, 2006astro.ph..8280K, 2015MNRAS.448..928K, 2017AcA....67...37C, 2019ApJ...870...71P, 2021MNRAS.507.5658G}, 
    star evolution \citep{2010ApJ...715L.138B, 2012ApJ...759...52D,2014AA...564A.134M,2019MNRAS.487.1675A, 2020AA...638A..94O, 2022MNRAS.509.1557C}, 
    short gamma-ray bursts \citep{2013ApJ...767..140P,2015ApJ...811L..22J}, 
    and gravitational wave events \citep{2019PhRvX...9c1040A,2020ApJ...892L...3A,2021ApJ...913L...7A}, respectively.
    The values from the gravitational wave events in Gpc$^{-3}$ yr$^{-1}$ are converted into the Myr$^{-1}$ unit under the assumption of a galactic density of 0.01 galaxy Mpc$^{-3}$.}
    \label{fig:NSMrate}
\end{figure}

The NSM rates from previous studies are summarized in Figure \ref{fig:NSMrate}. 
The NSM rates are estimated using several methods, and even though the values approach each other recently, they still have non-negligible uncertainties or systematic errors that are dependent on the methods used. 
According to Figure \ref{fig:plot_sensitivity},
instruments with higher sensitivities can cover more than 10\% of NSMs and should be able to observe multiple Galactic NSM remnants
(meaning a sensitivity of $10^{-9.5}$--$10^{-8.5}$ photons s$^{-1}$ cm$^{-2}$ or
$10^{-16.5}$--$10^{-14.5}$ erg s$^{-1}$ cm$^{-2}$ in the hard X-ray to $\gamma$-ray bands).
The actual numbers observed by future NSM surveys with highly sensitive instruments will provide direct information for the NSM rate in the local universe.

\subsection{Sensitivity requirements for future missions}
\label{sec:discussion:future}

\begin{figure}[hb]
    \centering
    \includegraphics[width=0.45 \textwidth]{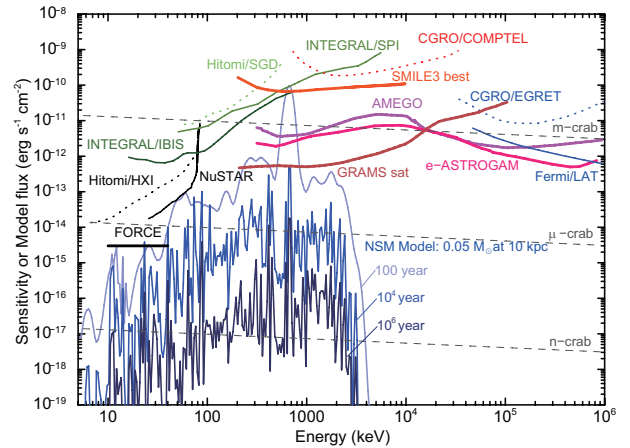}
    \caption{The $3 \sigma$ sensitivities of the missions in the hard X-ray to $\gamma$-ray bands are compared with the $\gamma$ rays expected from NSMs that are 100, $10^4$, and $10^6$ years old, with the assumption of an ejecta mass of 0.05 $M_\cdot$ at a 10 kpc distance with $\Delta E = 3$ keV.
    Past and future missions are shown with the dotted and thick lines, respectively.
    The sensitivities are taken from the following references: 
    CGRO/COMPTEL (9 years), INTEGRAL/SPI (1 year), e-ASTROGAM (1 year), and Fermi LAT (10 years in survey mode) 
    from \citet{2018JHEAp..19....1D}, 
    INTEGRAL/IBIS ($10^6$ sec) and AMEGO (5 years) from \citet{2020SPIE11444E..31K},
    CGRO/EGRET (9 years) from \citet{2021ChA&A..45..281S}, 
    NuSTAR ($10^6$ sec) from \citet{2013ApJ...770..103H}, 
    FORCE (3.5$\sigma$, $10^6$ sec) from \citet{2018SPIE10699E..2DN}, 
    Hitomi HXI and SGD (100 ksec) from \citet{2014SPIE.9144E..25T}, 
    SMILE3 ($10^6$ sec, best condition) from \citet{2021arXiv210700180T}, 
    and GRAMS (1 year) from \citet{2020APh...114..107A}.
    For reference, 1.0, $10^{-3}$, $10^{-6}$, and $10^{-9}$ times the hard X-ray flux from the Crab Nebula \citep{2013PASJ...65...74K}, with a simple extension into the $\gamma$-ray band with the single power-law spectrum, are shown in dashed lines and are noted as m-crab, $\mu$-crab, and n-crab, respectively.}
    \label{fig:sensitivities_missions}
\end{figure}

To assess the feasibility of detecting Galactic NSM remnants using past, current, and future $\gamma$-ray missions,
the $\gamma$-ray spectra expected from NSMs (Figures \ref{fig:rprocess_g_model_plot1} and \ref{fig:rprocess_g_model_plot2}) 
are compared with the sensitivities of the instruments for these missions in Figure \ref{fig:sensitivities_missions}.
For the MeV bands, we expect that in the 2030s sensitivities will be achieved that are one or two orders of magnitude higher than those of current missions.
Consequently, we conclude that future missions, such as e-ASTROGAM, AMEGO, and GRAMS, have the potential 
to detect MeV emissions from young NSM remnants in the age range of $t=$ 100--1000 years old
at 10 kpc, with a sensitivity of approximately $10^{-13}$--$10^{-11}$ erg cm$^{-2}$ s$^{-1}$.
Furthermore, the hard X-ray band below 100 keV is also useful in searching for NSM remnants. 
NuSTAR data may be able to indicate very young NSM remnants at about $t=$ 100 years,
and FORCE may be able to detect emissions from an NSM older than $t<10^4$ years at 10 kpc.

\acknowledgments
This work was supported in part by JSPS KAKENHI [Grant Nos. JP18H04571 and JP20K04009 (YT), 18H01232 and 22H01251(RY), 20K03957 (SF), 20H00174 (SK), 21H01121 (SK, YT, SF), 19K03908 (AB)]. 
YT and SK are deeply appreciative of the Observational Astrophysics Institute at Saitama University for supporting the research fund, and R.Y. deeply appreciates the Aoyama Gakuin University Research Institute for helping to fund our research.
Finally, we would like to thank the anonymous referee for
his/her careful reading of our manuscript and helpful comments.

%
\vspace{5mm}
\facilities{INTEGRAL, CGRO, NuSTAR, Hitomi, Fermi, SMILE, GRAMS, AMEGO, e-ASTROGAM, FORCE}


\bibliography{gammasim_yterada_apj}{}
\bibliographystyle{aasjournal}

\appendix
\section{XSPEC Model for gamma rays from r-process objects}
\label{sec:appendix:xspec}

The $\gamma$-ray spectra for NSM remnants in this paper are implemented as the table spectral model for XSPEC \citep{1996ASPC..101...17A} version 12 in the HEAsoft package. 
The model is provided with the file name of {\it rprocgamma.mod} in the flexible image transport system (FITS) format \citep{2001A&A...376..359H}.
The model parameters and descriptions are summarized below. 
The model does not contain the Doppler effect, but instead it gives the output of the second step described in Section \ref{sec:estimation:summary}.

\begin{itemize}
    \item {\bf time :} time from the merging event, in units of years.
    \item {\bf Ye$NN$ ($NN=10,15,20,25,30,35,40,45$):} the ejecta mass of the $r$-process nuclei in solar mass units $M_\odot$ under the $Y_{\rm e}=0.NN$ environment.
    \item {\bf $z$ :} redshift
    \item {\bf norm :} normalization in units of photons s$^{-1}$ cm$^{-2}$ $d^{-2}$ $(1+z)^{-2}$, where $d$ is the distance to the object in kpc.
    
\end{itemize}

For example, the spectral model of a Galactic NSM remnant at $t=100.0$ years and $d=10$ kpc,
calculated under the assumption that the ejecta masses of 
$Y_{\rm e}=$0.10, 0.15, 0.20, 0.25, 0.30, 0.35, 0.40, and 0.45 
are 0.0454, 0.0485, 0.146, 0.297, 0.103, 0.251, 0.105, and 0.00326 $M_\odot$, respectively, 
without the Doppler-broadening effect, is described with the following parameters:
\begin{verbatim}
Model atable{rprocgamma.mod}<1> Source No.: 1   Active/Off
Model Model Component  Parameter  Unit     Value
 par  comp
   1    1   rprocgamma time       year     100.000      +/-  0.0 
   2    1   rprocgamma Ye10       Msun     4.54000E-02  +/-  0.0
   3    1   rprocgamma Ye15       Msun     4.85000E-02  +/-  0.0
   4    1   rprocgamma Ye20       Msun     0.146000     +/-  0.0
   5    1   rprocgamma Ye25       Msun     0.297000     +/-  0.0
   6    1   rprocgamma Ye30       Msun     0.103000     +/-  0.0
   7    1   rprocgamma Ye35       Msun     0.251000     +/-  0.0
   8    1   rprocgamma Ye40       Msun     0.105000     +/-  0.0
   9    1   rprocgamma Ye45       Msun     3.26000E-03  +/-  0.0 
  10    1   rprocgamma z                   0.0          frozen
  11    1   rprocgamma norm                0.01000      +/-  0.0
\end{verbatim}

An example of the XSPEC commands to use this numerical model with the Doppler broadening effect are followings. The comments shown to the right of \%\% and are not executed.
For detail, please check the manual for XSPEC.
\begin{verbatim}
    XSPEC12> model gsmooth(atable{rprocgamma.mod})   %% set the model and parameters
    XSPEC12> cpd /xw              %% change the plot device using X-Windows system
    XSPEC12> dummyrsp 1 4000 1000 %% set a dummy response from 1 to 4,000 keV in 1,000 bins
    XSPEC12> plot model           %% plot the model on the screen
\end{verbatim}

\end{document}